\pdfoutput=1
\documentclass[prd,twocolumn,preprintnumbers,floatfix,
nofootinbib,superscriptaddress]{revtex4}
\usepackage{amsfonts}
\usepackage{bm,graphicx}
\usepackage{hyperref}

\pdfoutput=1 
\usepackage{amssymb,amsmath,hyperref}
\usepackage{color}
\usepackage{graphicx}
\usepackage{bm}
\usepackage{amsmath,amssymb}
\usepackage{graphicx}
\usepackage{bm}
\usepackage{comment} 
\usepackage{subfigure}
\usepackage{array}
\usepackage{multirow}

\newcommand{\nn}{\nonumber}

\newcommand{\Erho}{2.1762(28)-i0.0150(7)}
\newcommand{\mrho}{2.1762(28)}

\newcommand{\LG}{\mathrm{LG}}

\newcommand{\F}{F_{\pi\rho}}
\newcommand{\RLL}{\mathcal{R}}
\newcommand{\Kpipi}{\mathcal{K}_{\pi\pi}}
\newcommand{\Epipi}{E_{\pi\pi}}
\newcommand{\Amp}{\mathcal{A}_{\pi\pi,\pi\gamma^\star}}

\hypersetup{colorlinks, linkcolor = [rgb]{0,0.0,0.75}, citecolor = [rgb]{0,0.0,0.75}, urlcolor = [rgb]{0,0.0,0.75}}

\begin{document}

 \preprint{\vbox{
\hbox{JLAB-THY-16-2237} 
}}
 \preprint{\vbox{\hbox{DAMTP-2016-24 }}}


\title{The $\pi\pi\to\pi\gamma^\star$ amplitude and the resonant $\rho\to\pi\gamma^\star$ transition from lattice QCD}

\author{Ra\'ul A. Brice\~no}
\email[e-mail: ]{rbriceno@jlab.org}
\affiliation{Thomas Jefferson National Accelerator Facility, 12000 Jefferson Avenue, Newport News, VA 23606, USA}
\affiliation{Department of Physics, Old Dominion University, Norfolk, VA 23529, USA}
\author{Jozef J. Dudek}
\affiliation{Thomas Jefferson National Accelerator Facility, 12000 Jefferson Avenue, Newport News, VA 23606, USA}
\affiliation{Department of Physics, Old Dominion University, Norfolk, VA 23529, USA}
\author{Robert G. Edwards}
\affiliation{Thomas Jefferson National Accelerator Facility, 12000 Jefferson Avenue, Newport News, VA 23606, USA}
\author{Christian J. Shultz}
\affiliation{Department of Physics, Old Dominion University, Norfolk, VA 23529, USA}
\author{Christopher~E.~Thomas}
\affiliation{Department of Applied Mathematics and Theoretical Physics, Centre for Mathematical Sciences, University of Cambridge, Wilberforce Road, Cambridge, CB3 0WA, UK}
\author{David~J.~Wilson} 
\affiliation{Department of Applied Mathematics and Theoretical Physics, Centre for Mathematical Sciences, University of Cambridge, Wilberforce Road, Cambridge, CB3 0WA, UK}

\collaboration{for the Hadron Spectrum Collaboration}

\begin{abstract}
We present a determination of the $P$-wave $\pi\pi\to\pi\gamma^\star$ transition amplitude from lattice quantum chromodynamics. Matrix elements of the vector current in a finite-volume are extracted from three-point correlation functions, and from these we determine the infinite-volume amplitude using a generalization of the Lellouch-L\"uscher formalism. We determine the amplitude for a range of discrete values of the $\pi\pi$ energy and virtuality of the photon, and observe the expected dynamical enhancement due to the $\rho$ resonance. Describing the energy dependence of the amplitude, we are able to analytically continue into the complex energy plane and from the residue at the $\rho$ pole extract the $\rho\to \pi \gamma^\star$ transition form factor. This calculation, at $m_\pi\approx 400$~MeV, is the first to determine the form factor of an unstable hadron within a first principles approach to QCD.
 
\end{abstract}

\maketitle

\section{introduction \label{sec:intro}}

The study of hadron resonances is entering a new era: for the first time since the identification of quantum chromodynamics (QCD) as the fundamental theory of the strong interactions, one can realistically study resonances and their properties directly from QCD by taking advantage of numerical computations of the theory within the framework of lattice QCD.

Hadron resonances emerge as pole singularities in the scattering-matrix, or $S$-matrix, at complex values of the scattering energy. On the other hand, lattice QCD calculations being performed in a finite Euclidean volume results in a discrete real-valued spectrum, and this observation might lead one to conclude that resonances cannot be directly studied using lattice QCD. The way around this is to recognize that the spectrum of states in a finite-volume is determined by the infinite-volume $S$-matrix elements in a way that is known~\cite{Luscher:1986pf, Luscher:1990ux, Rummukainen:1995vs, Kim:2005gf, Christ:2005gi, Briceno:2012yi, Hansen:2012tf, Briceno:2014oea, Hansen:2014eka,  Hansen:2015zga, Hansen:2015zta, Polejaeva:2012ut,Briceno:2012rv}, so that knowledge of the discrete spectrum can lead to a determination of the $S$-matrix at real values of the energy. From this the extension to complex values of the energy can proceed, as in the experimental case, using parameterizations of the energy dependence analytically continued into the complex plane. The resonant structure follows from the pole singularities of the $S$-matrix. This methodology has been applied in order to determine the masses and widths of resonances that couple to two-body elastic~\cite{Dudek:2012xn,  Lang:2011mn, Lang:2015hza, Lang:2014yfa,  Feng:2010es, Pelissier:2012pi, Prelovsek:2013ela, Aoki:2011yj, Aoki:2007rd, Bolton:2015psa} and inelastic systems~\cite{Wilson:2015dqa, Wilson:2014cna, Dudek:2014qha, Wilson:2015a0}.

Hadron resonances can also appear in processes featuring electroweak currents, and recently the formalism required to study these in a finite-volume has been presented both for transitions~\cite{Briceno:2015csa, Briceno:2014uqa, Agadjanov:2014kha, Meyer:2011um} and elastic form factors~\cite{Briceno:2015tza, Bernard:2012bi}. These ideas generalize the existing framework for the study of $K\to\pi\pi$ decays, which was first proposed in the seminal work by Lellouch and L\"uscher~\cite{Lellouch:2000pv}, and whose numerical implementations have reached an impressive level of maturity~\cite{Bai:2015nea, Ishizuka:2014nfa, Blum:2012uk, Boyle:2012ys, Blum:2011pu, Blum:2011ng}. In this study we follow the procedure presented in Refs.~\cite{Briceno:2015csa, Briceno:2014uqa, Agadjanov:2014kha} to obtain the electromagnetic form factor of a hadronic resonance for the first time in QCD.

The quantity we determine is the $ \pi \gamma^\star \to\pi \pi $ amplitude, $\mathcal{H}_{\pi\pi,\pi\gamma^\star}^{\mu}$. To first order in QED interactions, this can be defined in terms of the electromagnetic current, ${   {\mathcal{J}}^{\mu}=\tfrac{2}{3}\bar{u}\gamma^\mu u- \tfrac{1}{3}\bar{d}\gamma^\mu d   }$, where $u$ and $d$ denote the annihilation up and down--quark fields
\footnote{The position space current is denoted as ${\mathcal{J}}^{\mu}(t,x)$, and its Fourier transform will be labeled as $\widetilde{\mathcal{J}}^{\mu}(t,\mathbf{Q})$. }
, as
\begin{align}
 \label{eq:infinite_volume_amp}
{\mathcal{H}_{\pi\pi,\pi\gamma^\star}^{\mu} }
=
{\big\langle \pi,P_\pi \big|{\mathcal{J}}^{\mu}(0)   \big| \pi\pi,P_{\pi\pi},\ell=1 \big\rangle}.
\end{align}
where $| \pi\pi,P_{\pi\pi},\ell=1 \rangle$ is an incoming $P$-wave $\pi\pi$ state with four-momentum $P_{\pi\pi}$ and  $\langle  \pi,P_\pi |$ is an outgoing $\pi$ state with four-momentum $P_\pi$. We will obtain this amplitude from corresponding finite-volume matrix elements computed using lattice QCD applying the non-perturbative mapping prescribed in Ref.~\cite{Briceno:2014uqa}. The amplitude is determined at a number of $\pi\pi$ energies and photon virtualities. Using these to constrain parameterizations of the $E_{\pi\pi}$ and $Q^2$ dependence, we analytically continue to the pole in the complex energy plane corresponding to the  $\rho$ resonance and obtain the residue of the amplitude, which contains the $\rho \to \pi \gamma^\star$ transition form factor. 

In addition to serving as a stepping stone towards the study of more complicated and computationally taxing resonant processes, $\pi\gamma^\star\to \pi\pi$ plays a significant role in the determination of various phenomenologically interesting observables. These include the anomalous magnetic moment of the muon~\cite{Colangelo:2014dfa, Colangelo:2014pva} and the Wess-Zumino-Witten anomaly~\cite{Wess:1971yu, Witten:1983tw} among others.

This first exploratory study is performed using a single value of degenerate $u,d$ quark masses, corresponding to ${m_\pi \approx 400\,\mathrm{MeV}}$. In this paper we expand upon the details of the calculation that appeared in summary form in Ref.~\cite{Briceno:2015dca}.  We make use of the technology laid out in Ref.~\cite{Shultz:2015pfa} for the computation of three-point correlation functions, and the results for the $\pi\pi$ elastic scattering phase shift determined from the lattice QCD spectrum in Ref.~\cite{Dudek:2012xn}. 
 
This work is presented as follows. In Sect.~\ref{sec:matrix_elements} we review the set up of the lattice calculation and the extraction of finite-volume matrix elements from correlation functions. We review the formalism needed to obtain the infinite-volume transition amplitude from the finite-volume matrix elements in Sect.~\ref{sec:FV}. Section~\ref{sec:global_fits} discusses the procedure used in fitting the transition amplitude and contains the main results of this work, the $\pi\pi \to \pi\gamma^\star$ transition amplitude and the $\rho \to \pi \gamma^\star$ form factor extracted at the $\rho$ pole. We present the ${\pi\gamma\to\pi\pi}$ cross section in Sect.~\ref{sec:cross_section} and then summarize the findings and implications of this work in Sect.~\ref{sec:conclusion}.

 \section{Three-point functions and matrix elements  \label{sec:matrix_elements}}

\begin{table}[t]
\subtable[]{ \label{table:ens_info}
  \begin{tabular}{c|ccc}
    $(L/a_s)^3 \times (T/a_t)$   &$N_{\mathrm{cfgs}}$ & $N_{\mathrm{t_{srcs}}}$ & $N_{\mathrm{vecs}}$ \\
    \hline 
     $20^3\times 128$ & 603 & 4 & 128\\   
  \end{tabular}
} \subtable[]{
\begin{tabular}{c|l}
$a_t m_\pi$    & $0.06906(13)$ \\
$a_t m_K$     & $0.09698(9)$ \\
$a_t m_\eta$   & $0.10406(56)$ \\  
$a_t m_\omega$  & $0.15678(41)$ \\ 
$a_t m_\Omega$  & $0.2951(22)$ \\ 
\hline
$\xi$ & $3.444(6)$
\end{tabular}
}
\caption{(a) The volume $((L/a_s)^3 \times (T/a_t))$, number of gauge configuration ($N_{\mathrm{cfgs}}$), number of sources ($N_{\mathrm{t_{srcs}}}$) and distillation vectors ($N_{\mathrm{vecs}}$) used in this calculation. (b) Some previously-determined low-lying hadron masses.}
\label{TABLE:ensemble}
\end{table}

The results presented in this calculation used an ensemble of gauge configurations with a Symanzik improved gauge action and a Clover fermion action with $N_f=2+1$ dynamical fermions. The quark masses are chosen so that $m_\pi\approx 400$~MeV ~\cite{Edwards:2008ja,Lin:2008pr}. We use a space-time volume of $(L/a_s)^3\,\times\,(T/a_t) = 20^3\times 128$, where the spatial lattice spacing is $a_s\approx 0.12$~fm and the temporal lattice spacing, $a_t$, is smaller with an anisotropy $\xi = a_s/a_t \approx 3.5$. We set the lattice scale using a procedure where $a_t = \tfrac{a_t m_\Omega}{m_\Omega^{\textrm{phys}}}$, using the $\Omega$ baryon mass determined on this lattice (see Table \ref{TABLE:ensemble}) and the physical $\Omega$ baryon mass.  The spatial and temporal extents, $m_\pi L\approx 4.7$ and $m_\pi T\approx 8.8$, are such that finite-volume and finite-temperature effects for single-hadron observables lie well below the percent level of precision and can be safely ignored, as demonstrated in Ref.~\cite{Dudek:2012gj}. This also ensures that all finite-volume corrections associated with the $\pi\pi\to\pi\gamma^\star$ matrix elements are those addressed in Refs.~\cite{Briceno:2015csa, Briceno:2014uqa} which are corrected nonperturbatively. We use the ``\emph{distillation}" technique~\cite{Peardon:2009gh} in the construction of both two-point and three-point correlation functions. Some details of the calculation and the size of the distillation basis, along with the masses of some low-lying hadrons, are summarized in Table~\ref{TABLE:ensemble}.

We can extract the desired matrix elements from three-point correlation functions of the form
\begin{widetext}
\begin{equation}
C^{(3)}_{\pi\pi,\mu, \pi}(\mathbf{P}_{\!\pi}, \mathbf{P}_{\!\pi\pi}; \Delta t, t) = \big\langle 0 \big| \mathcal{O}_\pi^{[\Lambda_\pi]}(\Delta t, \mathbf{P}_{\!\pi}) \; \widetilde{\mathcal{J}}_\mu(t, \mathbf{P}_{\!\pi} \!-\! \mathbf{P}_{\!\pi\pi} ) \; \mathcal{O}^{[\Lambda_{\pi\pi}]\dag}_{\pi\pi}(0, \mathbf{P}_{\!\pi\pi}) \big| 0 \big\rangle,
\end{equation}
\end{widetext}
where $ \widetilde{\mathcal{J}}_\mu(t, \mathbf{P}_{\!\pi} \!-\! \mathbf{P}_{\!\pi\pi} ) $ is the Fourier transform of the position-space current appearing in Eq.~\ref{eq:infinite_volume_amp}. In this expression $\mathcal{O}^{[\Lambda_{\pi}]}_{\pi}(\Delta t, \mathbf{P}_{\!\pi})$ is a composite QCD operator having the quantum numbers of a pion with three-momentum $\mathbf{P}_{\!\pi}$, evaluated at Euclidean time, $\Delta t$. The relevant irreducible representations, $\Lambda_{\pi}$, of the appropriate symmetry group are $A_1^+$ for a pion at rest and $A_2$ for a pion with any of the non-zero momenta we consider~\cite{Thomas:2011rh}. The operator $\mathcal{O}^{[\Lambda_{\pi\pi}]}_{\pi\pi}(0, \mathbf{P}_{\!\pi\pi})$ is constructed to have the quantum numbers of two pions with isospin=1 and total three-momentum $\mathbf{P}_{\!\pi\pi}$ in irreps $\Lambda_{\pi\pi}$ containing a subduction of the $\ell=1$ partial wave -- these irreps are listed in Table~\ref{table:irreps}. The vector current, $\widetilde{\mathcal{J}}_\mu(t, \mathbf{P}_{\!\pi} - \mathbf{P}_{\!\pi\pi} )$, is inserted at all times, $t$, between $0$ and $\Delta t$.

Time-evolving the operators and inserting complete sets of discrete finite-volume eigenstates of QCD leads to a spectral representation of the form
\begin{widetext}
\begin{equation}
\sum_{n,m} e^{- (\Delta t - t) E_{\pi, m} } \, e^{- t\, E_{\pi\pi, n} } \big\langle 0 \big| \mathcal{O}_\pi| \pi, m; L\big\rangle \, \big\langle \pi, m; L \big| \widetilde{\mathcal{J}}_\mu \big| \pi\pi, n; L \big\rangle \big\langle \pi\pi, n; L \big| \mathcal{O}_{\pi\pi}^\dag \big| 0 \big\rangle,
\end{equation}
\end{widetext}
which features contributions from transitions between all eigenstates with the correct quantum numbers. The finite-volume energy eigenstates which feature in this expression are defined with a normalization $\big\langle \pi\pi, n; L \big| \pi\pi, n; L \big\rangle = 1$ and obvious orthogonalities between different momenta and irreps (see Appendix~\ref{sec:convention}). For an arbitrary choice of operators, $\mathcal{O}_\pi, \mathcal{O}_{\pi\pi}$, this leads to pollution from excited states when trying to determine the ground-state transition, and it proves to be the case that excited states are not determined well by fitting their subleading time-dependence. A solution to this problem comes by using operators which optimally interpolate particular states in the spectrum, with minimal amplitude to produce any other state. Such operators can be constructed as linear superpositions in a basis of operators by `diagonalising' a matrix of \emph{two-point} correlation functions,
\begin{equation}
C^{(2)}_{ab}(t) = \big\langle 0 \big| \mathcal{O}_a(t) \, \mathcal{O}^\dag_b(0) \big| 0 \big\rangle.
\end{equation}
Solving the generalized eigenvalue problem, ${C(t) v_n = \lambda_n(t) C(t_0) v_n}$, the operator which optimally produces state $n$ can be constructed as
\begin{equation}
\Omega_n^\dag = e^{-\frac{1}{2} E_n t_0} \sum\nolimits_a (v_n)_a \, \mathcal{O}_a^\dag .
\end{equation}

These operators can be used in the construction of the relevant three-point functions to isolate the contributions of particular states. This technique was previously explored in Ref.~\cite{Shultz:2015pfa} for the case of transitions between stable single-meson states with pseudoscalar and vector quantum numbers, where it was found to reduce excited state contributions to the ground-state transitions and to allow access to excited state transitions.

A basis of operators appropriate to form an optimized operator for a single pion can be constructed from quark bilinears with gauge-covariant derivatives, $\bar{q} \Gamma \overleftrightarrow{D} \ldots \overleftrightarrow{D} q$ -- what we will refer to as ``$\bar{q}q$-like'' operators, as was previously explored in Refs.\cite{Dudek:2009qf, Dudek:2010wm, Dudek:2011tt, Thomas:2011rh, Liu:2012ze, Moir:2013ub, Dudek:2013yja, Dudek:2012gj, Dudek:2010ew}. In the case of operators with the quantum numbers of two pions, in Refs.~\cite{Dudek:2012xn,Wilson:2015dqa} it was found that the corresponding discrete spectrum of states can be efficiently obtained using a basis of operators including both constructions built from the product of two optimal pion operators, $ \sum_{\hat{\mathbf{P}}_1, \hat{\mathbf{P}}_2} C(\mathbf{P}_{\!\pi\pi}; \mathbf{P}_1, \mathbf{P}_2)\, \Omega_{\pi}(t, \mathbf{P}_{1})\, \Omega_{\pi}(t, \mathbf{P}_{2})$, and ``$\bar{q}q$-like'' operators with the appropriate quantum numbers. The optimized operators in this channel prove to be superpositions featuring both forms.

Using optimized operators in three-point functions,
\begin{widetext}
\begin{align}
C^{(3)}_{\pi\pi_n,\mu, \pi}(\mathbf{P}_{\!\pi}, \mathbf{P}_{\!\pi\pi}; \Delta t, t) &= \big\langle 0 \big| \Omega_\pi^{[\Lambda_\pi]}(\Delta t, \mathbf{P}_{\!\pi}) \; \widetilde{\mathcal{J}}_\mu(t, \mathbf{P}_{\!\pi} - \mathbf{P}_{\!\pi\pi} ) \; \Omega^{[\Lambda_{\pi\pi, n}]\dag}_{\pi\pi}(0, \mathbf{P}_{\!\pi\pi}) \big| 0 \big\rangle \nn \\
&= e^{- (E_{\pi\pi, n} - E_\pi) t } \, e^{- E_\pi \Delta t} \,  \big\langle \pi; L \big| \widetilde{\mathcal{J}}_\mu \big| \pi\pi,n ; L \big\rangle + \ldots,
\label{eq:Cpipi_to_pi}
\end{align}
\end{widetext}
where the ellipsis should feature only modest contributions from states other than the single pion and the selected $n^\mathrm{th}$ $\pi\pi$ state. The optimal operators are constructed as linear superpositions in the basis outlined in Table~\ref{table:irreps}, and further details can be found in Ref.~\cite{Dudek:2012xn}.

Just as the operators, the finite volume states depend on the momentum of the system and irrep of the corresponding symmetry group, but we have suppressed these dependencies above. To avoid notational clutter, in the remainder of the text we highlight the dependencies of the states that play an important role in the subsequent equations. Given that we are only interested in the ground state with the quantum numbers of the $\pi$, we have dropped any labels which indicate so. Similarly, in the following discussion it will always be evident which $\pi\pi$ state is under consideration, and as a result, we will remove the label ``$n$''.

\begin{figure*}
\begin{center}
\includegraphics[width = \textwidth]{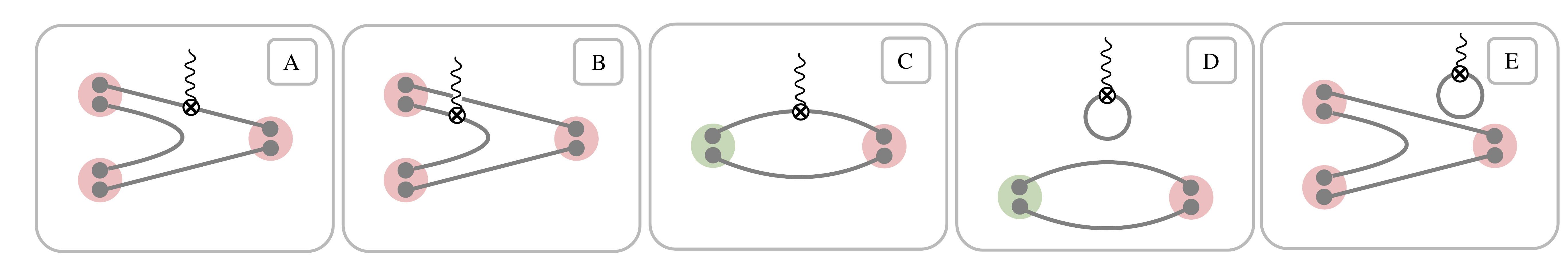}
\caption{Wick contractions that appear in the evaluation of three-point functions, $C^{(3)}_{\pi\pi,\mu, \pi}$, defined in Eq.~(\ref{eq:Cpipi_to_pi}). In this work we do not evaluate types D and E which feature a disconnected current insertion. }
\label{fig:contractions}
\end{center}
\end{figure*}

In order to compute these three-point correlation functions it is necessary to combine quark propagators in the arrangements shown in Figure~\ref{fig:contractions}. While we evaluate the diagrams of type A, B, and C, we set equal to zero the contribution of ``disconnected current'' diagrams of types D and E. These diagrams, which feature quark propagation to and from all points on the lattice, are computationally costly, and in the case we are considering we expect them to make only a small contribution. At the $SU(3)_F$ point, where up, down and strange quarks are mass degenerate, these contributions exactly cancel~\cite{Shultz:2015pfa}, and there are phenomenological reasons to expect that they do not become large as we reduce the light quark mass down from this point.

\begin{figure}
\begin{center}
\includegraphics[width = \columnwidth]{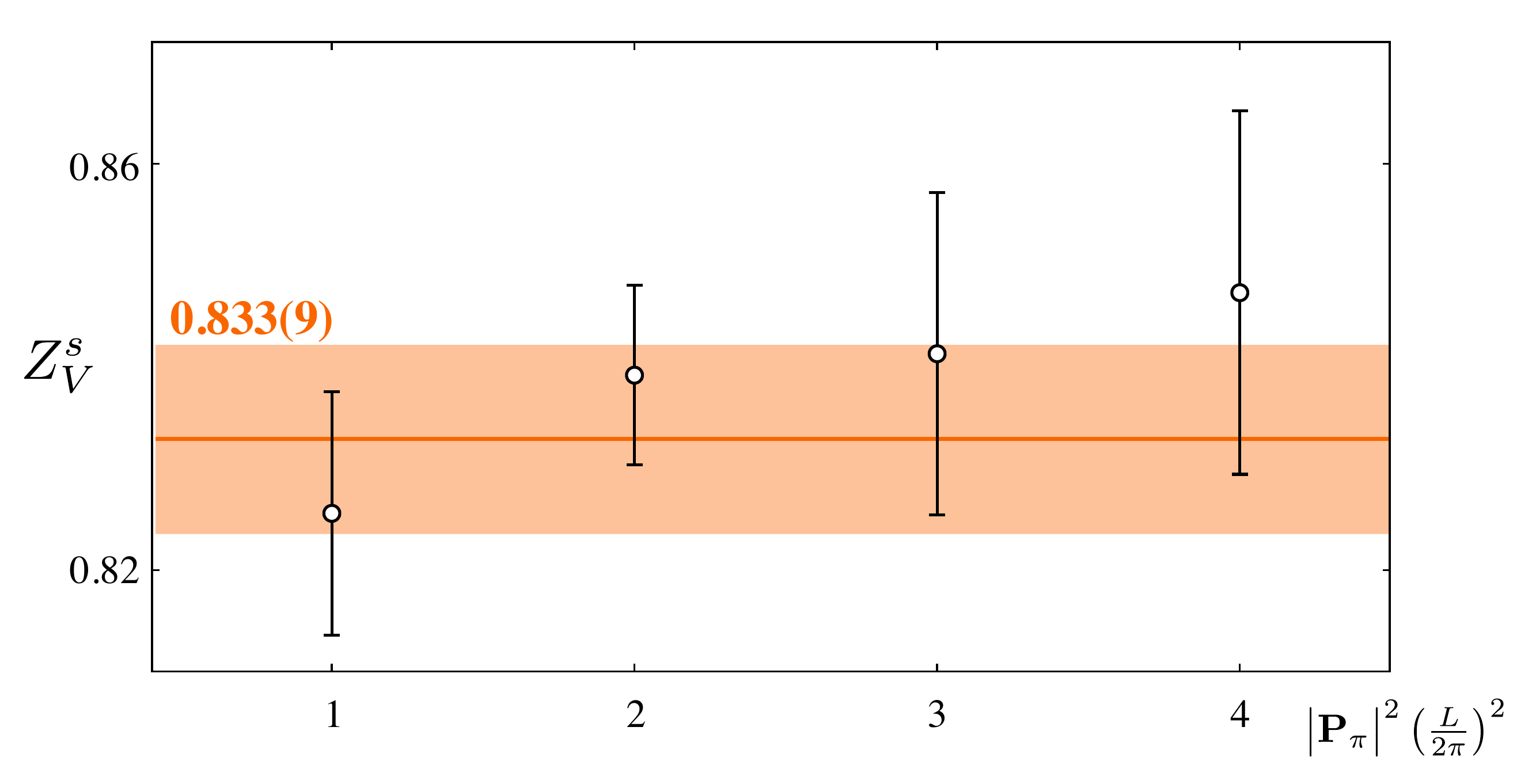}
\caption{Inverse of the unrenormalised $\pi$ form factor at $Q^2=0$, extracted using a spatially directed current insertion, as a function of the momentum of the source and sink pion. This corresponds to the vector current renormalization factor, $Z^s_V$. }
\label{fig:ZVs}
\end{center}
\end{figure}

The correlation functions are computed using the spatial component of the vector current; we use the tree-level improved Euclidean current to remove $\mathcal{O}(a)$ discretization effects on our anisotropic lattice~\cite{Shultz:2015pfa},
\begin{align}
\widetilde{\mathcal{J}}_{k} &= Z_V^s \Big(   \bar{q}\gamma_k q + \tfrac{1}{4} ( 1 - a_s/a_t) \, a_t \partial_4 \big( \bar{q}\sigma_{4k} q\big) \Big),
\end{align}
where $\gamma_k$ are the standard Euclidean-space gamma-matrices and $\sigma_{4k}=i[\gamma_4,\gamma_k]/2$. The vector current renormalization factor, $Z^s_V=0.833(9)$, is determined nonperturbatively by requiring the $\pi$ form factor, $F_\pi(Q^2)$, to be equal to one at $Q^2=0$. Fig.~\ref{fig:ZVs} shows unrenormalised values of the inverse of the form factor at four values of $\textbf{P}_{\!\pi}$, along with an appropriate average that leads to our value of $Z_V^s$.

\begin{table}[tb]
\begin{ruledtabular}
\begin{tabular}{cc|cccl }
$\textbf{P}$ & $\LG(\textbf{P})$ & State& $\Lambda^{(P)}$ & Operators\\
\hline \hline
\multirow{2}{*}{$[0,0,0]$} & \multirow{2}{*}{$\text{O}^{\text{D}}_h$} 
   &$\pi$ & $A_1^-$ & 12~$``\bar{q}q"$\\
   &   &$\pi\pi$& $T_1^-$ & $2~``\pi\pi",$ $26~``\bar{q}q"$\\
   \hline
\multirow{3}{*}{$[0,0,1]$} & \multirow{3}{*}{$\text{Dic}_4$} 
      &$\pi$& $A_2$ & 20~$``\bar{q}q"$\\
 &     &$\pi\pi$ & $A_1$ & $3~``\pi\pi",$ $27~``\bar{q}q"$\\
 &    &$\pi\pi$& $E_2$ & $2~``\pi\pi",$ $29~``\bar{q}q"$\\
\hline
\multirow{4}{*}{$[0,1,1]$} & \multirow{4}{*}{$\text{Dic}_2$} 
     &$\pi$ & $A_2$ & 31~$``\bar{q}q"$\\
 &    &$\pi\pi$  & $A_1$ & $3~``\pi\pi",$ $27~``\bar{q}q"$\\
 &   &$\pi\pi$ & $B_1$ & $3~``\pi\pi",$ $28~``\bar{q}q"$\\
\hline
\multirow{3}{*}{$[1,1,1]$} & \multirow{3}{*}{$\text{Dic}_3$} 
      &$\pi$& $A_2$ &21~ $``\bar{q}q"$\\
 &    &$\pi\pi$  & $A_1$ & $3~``\pi\pi",$ $21~``\bar{q}q"$\\
 &    &$\pi\pi$& $E_2$ & $2~``\pi\pi",$ $35~``\bar{q}q"$\\
\end{tabular}
\end{ruledtabular}
\caption{The momenta, $\textbf{P}$ (given in units of $2\pi/L$), with corresponding symmetry groups, $\LG(\textbf{P})$, and irreps, $\Lambda^{(P)}$, used to study the $\pi$ and $\pi\pi$ finite-volume states. For each irrep, the numbers of ``$\bar{q}q$''-like fermion bilinear and ``$\pi\pi$''-like operators used to construct optimal operators are shown. In the case of $\pi\pi$ we consider only those irreps which feature a subduction of $\ell=1$. Further details appear in Refs.~\cite{Thomas:2011rh, Dudek:2012xn}.}
\label{table:irreps}
\end{table}

In Ref.~\cite{Dudek:2012xn} it was demonstrated that $I=1$ $\pi\pi$ elastic scattering below $K\overline{K}$ threshold is dominated by the \mbox{$P$-wave} where the $\rho$ resonance resides, and as a result, it is expected that the $\pi\pi\to\pi\gamma^\star$ process in this energy region will be dominated by the $\ell=1$ contribution. The infinite-volume matrix element $\big\langle \pi, \mathbf{P}_{\!\pi} \big| \mathcal{J}^\mu(0) \big| \pi\pi, \mathbf{P}_{\!\pi\pi} \big\rangle$ with the $\pi\pi$ system having $\ell=1$, can be Lorentz decomposed in the following way,
\begin{align}
\big\langle& \pi, \mathbf{P}_{\!\pi} \big| \mathcal{J}^\mu(0) \big| \pi\pi, \mathbf{P}_{\!\pi\pi} \big\rangle  \nn \\
&= \epsilon^{\mu \nu \rho \sigma} (P_{\pi})_\nu\,  (P_{\pi\pi})_\rho\,  \epsilon_\sigma(\lambda_{\pi\pi}, \mathbf{P}_{\!\pi\pi})  \, \frac{2}{m_\pi} \mathcal{A}_{\pi\pi, \pi \gamma^\star}(E^\star_{\pi\pi}, Q^2),
\label{eq:decomp}
\end{align}
where $\epsilon_\sigma(\lambda_{\pi\pi}, \mathbf{P}_{\!\pi\pi})$ is a polarization vector describing the $\ell=1$ $\pi\pi$ system with helicity $\lambda_{\pi\pi}$, and $\mathcal{A}_{\pi\pi, \pi \gamma^\star}(E^\star_{\pi\pi}, Q^2)$ is a reduced amplitude depending upon the $\pi\pi$ cm-frame energy and the virtuality of the photon, ${Q^2 = -(P_{\pi\pi} - P_\pi)^2}$. In Appendix~\ref{sec:covariant_relation} we show that this decomposition of the transition amplitude is equivalent to another commonly used form.

Infinite-volume one-hadron states have the standard relativistic normalization (see Eq.~\ref{eq:infty_states}) and have dimensions of $\rm [MeV]^{-1}$. Two-hadron states constructed as products of two one-hadron states, have dimensions of $\rm [MeV]^{-2}$, and in position space, the current has units of $\rm [MeV]^{-3}$. Thus the left-hand side of Eq.~\ref{eq:decomp} and $\mathcal{A}_{\pi\pi, \pi \gamma^\star}$ have dimensions of $\rm [MeV]^{0}$ and $\rm [MeV]^{-1}$, respectively. 

A reasonable extension of the above decomposition to the $L\times L \times L$ finite-volume case is
\begin{align}
\big\langle& \pi, \mathbf{P}_{\!\pi}; L \big| \mathcal{J}^\mu(0) \big| \pi\pi, \mathbf{P}_{\!\pi\pi}; L \big\rangle \nonumber \\
&=
\frac{1}{L^3}
\big\langle \pi, \mathbf{P}_{\!\pi}; L \big| \widetilde{\mathcal{J}}^\mu(0,\mathbf{P}_{\!\pi} \!-\! \mathbf{P}_{\!\pi\pi}) \big| \pi\pi, \mathbf{P}_{\!\pi\pi}; L \big\rangle \nn
\\
&= \tfrac{1}{\sqrt{4 E_\pi E_{\pi\pi} } } \tfrac{1}{L^3} \,  \epsilon^{\mu \nu \rho \sigma} (P_{\pi})_\nu\,  (P_{\pi\pi})_\rho\,  \epsilon_\sigma(\lambda_{\pi\pi}, \mathbf{P}_{\!\pi\pi})  \nonumber \\ 
&\quad\quad\quad\quad \times \frac{2}{m_\pi} \tilde{\mathcal{A}}(E^\star_{\pi\pi}, Q^2; L),
\label{eq:mat_covariant}
\end{align}
where we have allowed the reduced amplitude, $\tilde{\mathcal{A}}(E^\star_{\pi\pi}, Q^2; L)$, to be volume dependent. In Appendix~\ref{sec:F_wave}, we discuss the implications of neglecting contributions due to partial waves higher than $\ell=1$. 

Performing a similar dimensional analysis as above and recognizing that one- and two-particle finite-volume states are unit-normalized, one finds that $\tilde{\mathcal{A}}$ is dimensionless. The precise relationship between the quantity we can extract from finite-volume three-point functions, $\tilde{\mathcal{A}}(E^\star_{\pi\pi}, Q^2; L)$, and the desired infinite-volume quantity, $\mathcal{A}_{\pi\pi, \pi \gamma^\star}(E^\star_{\pi\pi}, Q^2)$, will be described in Section~\ref{sec:trans_amp}, where it will be shown to depend upon the elastic $\pi\pi$ scattering amplitude.

\begin{figure}
\begin{center}
\includegraphics[width = \columnwidth]{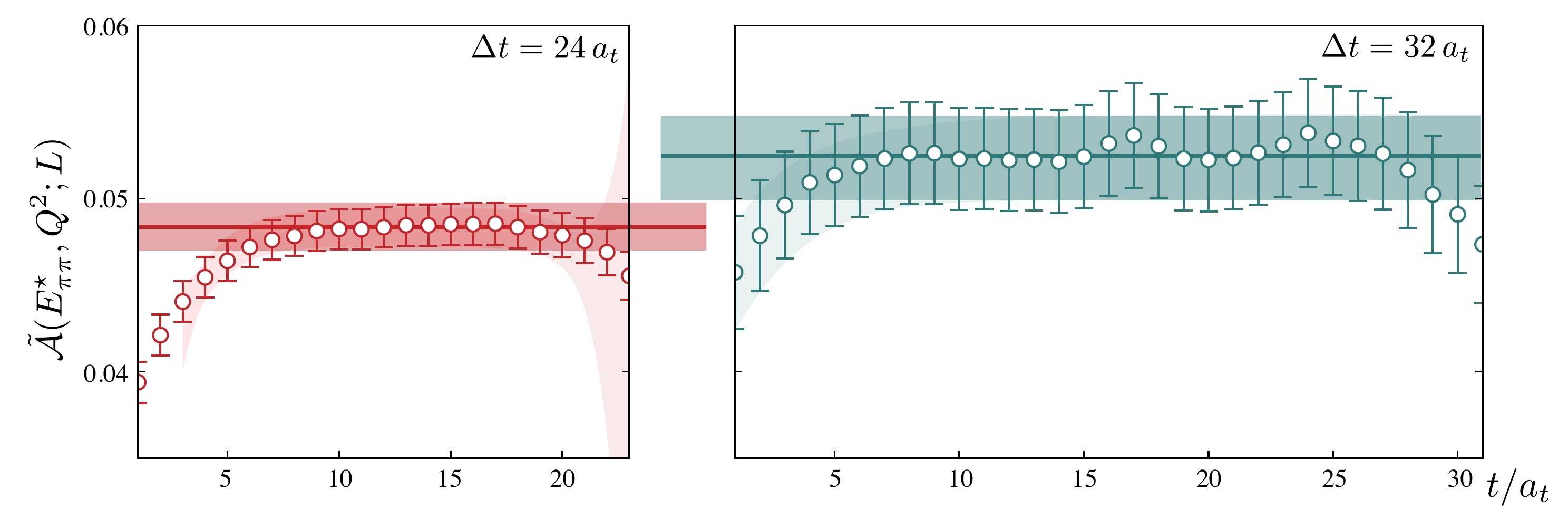}
\caption{Example of matrix elements determined from three-point correlators, as described in the text, with source-sink separations $ \Delta t/a_t=24$ (left) and $\Delta t/a_t=32$ (right). Correlated fits to the time-dependence give values that are statistically compatible.}
\label{fig:comparison_plot}
\end{center}
\end{figure}

Three-point functions were evaluated with two different time separations between source and sink operators, $\Delta t = 24 a_t$ and $32 a_t$.  Figure~\ref{fig:comparison_plot} illustrates an example of the matrix elements that we obtain on each timeslice after dividing out the leading exponential time-dependence in Eq.~\ref{eq:Cpipi_to_pi} and the kinematic prefactor in Eq.~\ref{eq:mat_covariant}. There are clearly plateau regions for both $\Delta t$. We fit the time-dependence using a form $a + b \, e^{-\delta E_1 (\Delta t - t)} + c\,  e^{-\delta E_2 t}$ that allows for residual excited state contributions from source and sink, and then $a$ gives the extracted value of $\tilde{\mathcal{A}}(E^\star_{\pi\pi}, Q^2; L)$.
We find for all our matrix elements that the results for the two time separations are statistically compatible and in what follows we conservatively choose to use the $\Delta t = 32 a_t$ results with their larger statistical uncertainties.

\begin{figure*}[h]
\begin{center}
\includegraphics[width = 0.8\textwidth]{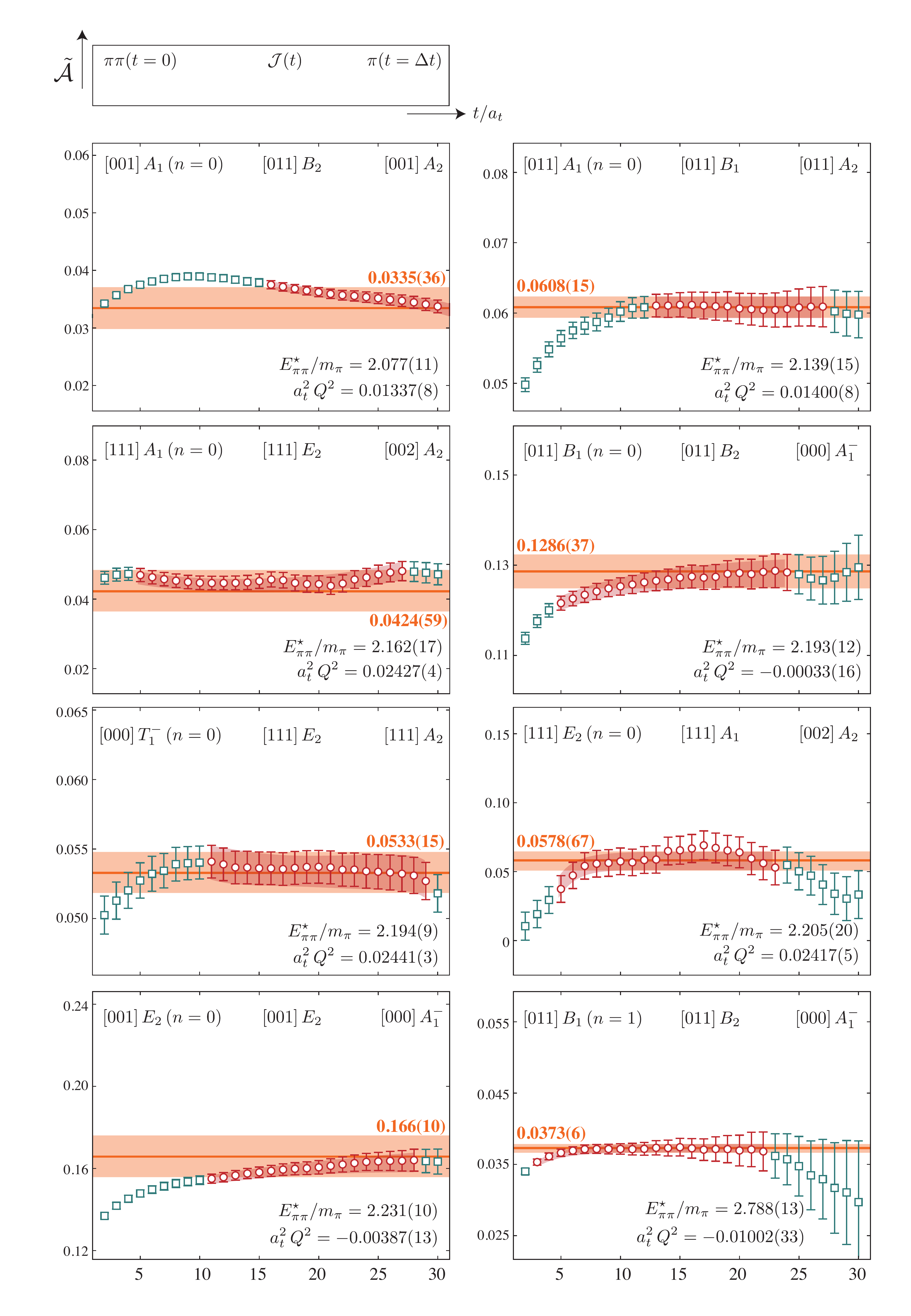}
\caption{Each panel shows the extracted matrix element as a function of time from a particular level in a $\pi\pi$ irrep. The red circles show the points used in the fit of time-dependence described in the text, while the blue points are not used. 
The red band is the time-dependent fit, the orange line and band show the central value of $\tilde{\mathcal{A}}$ extracted from the fit and one standard deviation on either side.  The label for each panel indicates, from left to right, the momentum and irrep of the $\pi\pi$ operator, the current insertion (subduced into an irrep, see \cite{Shultz:2015pfa}) and the $\pi$ operator.
 }
\label{fig:Fpi_tilde_abs}
\end{center}
\end{figure*}

We computed around $500$ matrix elements with various combinations of $\mathbf{P}_{\!\pi}$, $\mathbf{P}_{\!\pi\pi}$, irrep rows, and insertion direction, and from combinations of these we obtain 42 independent non-zero values of $\tilde{\mathcal{A}}(\Epipi^\star,Q^2,L)$ corresponding to 8 $\pi\pi$ energies and a range of $Q^2$ between $- 3m_\pi^2$ and $+7m_\pi^2$. In Fig.~\ref{fig:Fpi_tilde_abs} we give one example for each $\pi\pi$ irrep to illustrate the statistical quality of the determined matrix elements. The bottom right panel of Fig.~\ref{fig:Fpi_tilde_abs} corresponds to the \emph{first excited} state in the $B_1$ irrep with $\mathbf{P}_{\!\pi\pi} = [011]$. This extraction is made possible by the use of an operator optimized to overlap with the first-excited state. In all cases more residual excited state contribution is seen to arise from the $\pi\pi$ source at $t=0$ than from the $\pi$ source at $t=\Delta t$, but both are seen to be modest and can be described using subleading exponentials in a fit to the time-dependence.

\begin{center}
\begin{table} 
\begin{tabular}{l|l|l} 
 $\alpha_{20,A_1}^{[00n]} =\frac{2}{\sqrt{5}}$
 & $\alpha_{20,A_1}^{[nn0]} =-\frac{1}{\sqrt{5}}$
 & $\alpha_{22,A_1}^{[nnn]} =-2i\sqrt{\frac{6}{5}}$
 \\ 
 $\alpha_{20,E_2}^{[00n]} =-\frac{1}{\sqrt{5}}$
  & $\alpha_{22,A_1}^{[nn0]} =-i\sqrt{\frac{6}{5}}$
 & $\alpha_{22, E_2}^{[nnn]}  =i\sqrt{\frac{6}{5}}$
 \\
 & $\alpha_{20,B_1}^{[nn0]} =-\frac{1}{\sqrt{5}}$\\
&$  \alpha_{22,B_1}^{[nn0]} =i\sqrt{\frac{6}{5}}$\\
& $\alpha_{20,B_2}^{[nn0]} =\frac{2}{\sqrt{5}}$
\end{tabular}
\caption{Nonzero values of $\alpha_{20,\Lambda}^\mathbf{P}$ and $\alpha_{22,\Lambda}^\mathbf{P}$, featuring in the expression for the pseudo-phase, Eq.~(\ref{eq:Ppseudo-phase}).   
}
\label{table:alphad}
\end{table}
\end{center} 

\section{Relating finite and infinite volume quantities\label{sec:FV}}

Having obtained the discrete spectrum of states and transition matrix elements in a finite volume, our task is to obtain the corresponding infinite volume scattering and transition amplitudes. The extraction of the $\pi\pi$ $P$-wave elastic scattering amplitude, expressed in terms of the phase shift, $\delta_1(E_{\pi \pi}^\star)$, from the spectrum information was carried out in Ref.~\cite{Dudek:2012xn}, and we briefly summarize the method here.

\subsection{The $\pi\pi$ spectrum and the $P$-wave scattering phase shift}

For energy levels above the lowest two-particle threshold, but below the lowest relevant three or four-particle threshold, there exists a relation between the finite-volume spectrum and the infinite-volume scattering amplitudes, $\mathcal{M}$,~\cite{Luscher:1986pf, Luscher:1990ux, Rummukainen:1995vs, Kim:2005gf, Christ:2005gi}, that may be written,
\begin{equation}
\label{eq:QC_master}
\det[F^{-1}(P,L) + \mathcal M(P)] = 0\,,
\end{equation}
where $F^{-1}(P,L)$ is a function which in general depends on the geometry and size of the spatially periodic volume, and the two-particle four-momentum, $P$. Both $F$ and $\mathcal M$ are matrices in the space of partial-waves $\ell$ and of open scattering channels, and the determinant is evaluated over this space. The $\ell$ values which feature are those subduced into the relevant irrep of the reduced rotational symmetry group. Having obtained the finite-volume spectrum from lattice QCD computation, $F(P,L)$ is determined, which in turn allows one to constrain the scattering matrix. 
 
For sufficiently low energies, partial waves above the lowest one appearing in the relevant irrep are expected to be kinematically suppressed by the angular momentum barrier at threshold which ensures that ${\mathcal{M}_\ell = \frac{16 \pi }{\rho(E^\star)} \frac{1}{\cot \delta_\ell - i} \sim q^{\star2\ell}}$, where the phase-space ${\rho(E^\star) = 2\, q^\star / E^\star }$.

For the isotriplet $\pi\pi$ system below the $K\overline{K}$ threshold, we expect the scattering amplitude to be dominated by the $\ell=1$ channel, where the $\rho$-resonance resides, with contributions to the spectrum from $\ell\geq3$ partial waves being negligible (and indeed this was shown explicitly in Ref.~\cite{Dudek:2012xn}). In this case the determinant condition above reduces to a simple one-to-one mapping between the spectrum and the \mbox{$P$-wave} scattering phase shift, $\delta_{1}(E_{\pi\pi}^\star)$,
\begin{align}
\cot\delta_{1}(E_{\pi \pi}^\star)  +   \cot\phi^{\mathbf{P}, \Lambda}(E_{\pi \pi}^\star)   &=0,
\label{eq:QC}
\end{align}
where the pseudo-phase factor $\cot\phi^{\mathbf{P}, \Lambda}(E_{\pi \pi}^\star)$ is given by
\begin{align}
\cot\phi^{\mathbf{P}, \Lambda}(E_{\pi \pi}^\star) & \equiv \cot\phi^{\mathbf{P}}_{00}
+\alpha_{20,\Lambda}^\mathbf{P}\cot\phi^{\mathbf{P}}_{20}
+\alpha_{22,\Lambda}^\mathbf{P}\cot\phi^{\mathbf{P}}_{22},
\label{eq:Ppseudo-phase}
\end{align}
and the constants $\alpha_{2m,\Lambda}^\mathbf{P}$ are presented in Ref.~\cite{Briceno:2014uqa} and reproduced in Table~\ref{table:alphad}. We have introduced the functions $\phi^{\mathbf{P}}_{\ell m}(E_{\pi \pi}^\star)$, which can be written in terms of the generalized Zeta functions (see e.g. \cite{Rummukainen:1995vs}),
\begin{align}
\cot\phi^{\mathbf{P}}_{\ell m}=-\frac{(4\pi)^{3/2}}{q_{\pi\pi}^{\star\ell +1} \, \gamma \, L^3} \left(\frac{2\pi}{L}\right)^{\ell -2}\mathcal{Z}^\mathbf{P}_{\ell m}\left[1;(q_{\pi\pi}^\star {L}/2\pi)^2\right].~~~
\label{eq:philm}
\end{align}

\begin{figure}
\begin{center}
\includegraphics[width = \columnwidth]{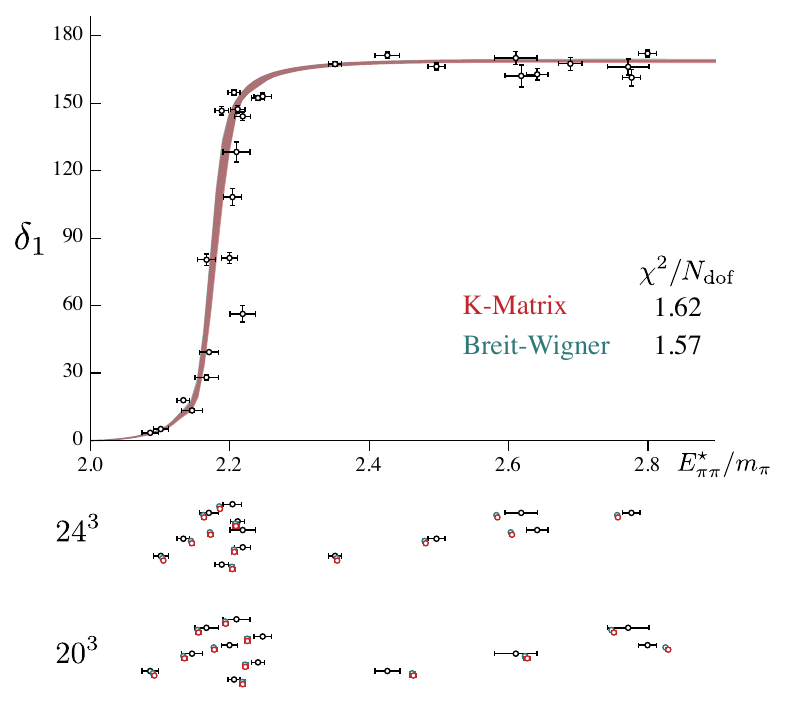}
\caption{Phase-shift values determined with Eq.~\ref{eq:QC} using energy levels from $20^3$ and $24^3$ lattices~\cite{Dudek:2012xn}. Parameterized descriptions using Breit-Wigner (Eq.~\ref{eq:bw}) and $K$-matrix (Eq.~\ref{eq:kmat}) forms also shown. The lower panels show the corresponding description of the finite-volume energy levels (black points) predicted using Breit-Wigner (blue) and $K$-matrix (red) parametrizations of the scattering phase shift. }
\label{fig:spectrum}
\end{center}
\end{figure}

In Figure~\ref{fig:spectrum} we show the phase-shifts which result from application of Eq.~\ref{eq:QC} to the finite-volume spectra obtained from $20^3$ and $24^3$ lattices~\cite{Dudek:2012xn}\footnote{For the current study the relevant two-point correlation functions were analysed independently with respect to Ref.~\cite{Dudek:2012xn}, and in some cases changes in choice of operator basis, choice of $t_0$, etc, led to a spectrum that is not identical to that presented in Ref.~\cite{Dudek:2012xn}. However, all determined levels agree up to shifts at the level of statistical fluctuations.}. A clear resonant behavior is observed, and two parameterizations of the elastic scattering amplitude which describe this spectrum well are the relativistic elastic Breit-Wigner,
\begin{align}
 \tan \delta_1(E_{\pi \pi}^\star) &= \frac{E_{\pi \pi}^\star \, \Gamma_{\mathrm{BW}}(E_{\pi \pi}^\star) }{m_\mathrm{BW}^2 - E_{\pi \pi}^{\star 2}}, \nonumber \\
 \Gamma_{\mathrm{BW}}(E_{\pi \pi}^\star) &= \frac{g_\mathrm{BW}^2 }{6\pi} \frac{q_{\pi\pi}^{\star 3} }{E_{\pi \pi}^2 },
 \label{eq:bw}
\end{align}
with parameters $m_\mathrm{BW}/m_\pi = 2.1780(29)$, $g_\mathrm{BW} = 5.82(8)$ and parameter correlation $+0.7$, and a single-channel Chew-Mandelstam $K$-matrix pole form,
\begin{align}
 \tan \delta_1(E_{\pi \pi}^\star) &= \frac{E_{\pi \pi}^\star \, \Gamma_{\mathrm{KM}}(E_{\pi \pi}^\star) }{m_\mathrm{KM}^2 - E_{\pi \pi}^{\star 2} + g_\mathrm{KM}^2 \, \delta I(E_{\pi \pi}^\star)      }, \nonumber \\
 \Gamma_{\mathrm{KM}}(E_{\pi \pi}^\star) &= {8\,g_\mathrm{KM}^2 } \frac{q_{\pi\pi}^{\star 3} }{E_{\pi \pi}^2 }, \nonumber \\
 \delta I(E_{\pi \pi}^\star)  &= \frac{\rho(E_{\pi \pi}^\star)}{\pi} \log \left[ \frac{\rho(E_{\pi \pi}^\star) + 1}{\rho(E_{\pi \pi}^\star) - 1} \right]   \nonumber \\
 &\quad - \frac{\rho(m_\mathrm{KM})}{\pi} \log \left[ \frac{\rho(m_\mathrm{KM}) + 1}{\rho(m_\mathrm{KM}) - 1} \right],
  \label{eq:kmat}
\end{align}
with parameters $m_\mathrm{KM}/m_\pi = 2.1790(39)$, $g_\mathrm{KM} = 0.465(8)$ and parameter correlation $-0.04$.

\subsection{Transition amplitude \label{sec:trans_amp}}

The process we are considering, $\pi \pi \to \pi \gamma^\star$, is an example of a ``$\mathbf{2} \to \mathbf{1}$'' transition induced by the vector current. The relationship between a finite-volume $\mathbf{2} \to \mathbf{1}$ matrix element and an infinite-volume transition amplitude was first given by Lellouch and L\"uscher~\cite{Lellouch:2000pv} for the case of $K\to\pi\pi$ decays induced by the weak current, where only the $\pi\pi$ $S$-wave could contribute. In our case, $\pi\pi\to\pi\gamma^\star$, the infinite-volume transition amplitude exists for many partial waves -- with $\pi\pi$ having $I=1$, all odd values of $\ell$ exist. 

As was the case for the spectrum, the reduced rotational symmetry of the cubic volume leads to infinitely many partial waves featuring in the relation between finite-volume matrix elements and infinite-volume transition amplitudes. This was first pointed out by Meyer in the context of bound state photodisintegration~\cite{Meyer:2012wk}, and later revisited for generic $\mathbf{2} \to \mathbf{1}$ transitions in Refs.~\cite{Briceno:2014uqa, Briceno:2015csa}, where it was shown that one can write a relation between a generic finite-volume  matrix element, ${\langle  \mathbf{1};L|\mathcal{J}^\mu(0)| \mathbf{2};L\rangle}$, and the corresponding infinite-volume transition amplitude, $\mathcal{H}^\mu_{\mathbf{2},\mathbf{1}} = \langle  \mathbf{1} |\mathcal{J}^\mu(0)| \mathbf{2} \rangle $. This relationship can be written\footnote{A factor of $L^3$ difference between what appears here and what is presented in Ref.~\cite{Briceno:2014uqa} is due to the fact that we are defining here the vector current in position space, rather than in momentum space, as was done there.}
\begin{align}
 \label{eq:master_equation}
\Big|{\langle  \mathbf{1};L|\mathcal{J}^\mu(0)| \mathbf{2};L\rangle}\Big|
=\frac{\sqrt{\big(\mathcal{H}^\mu_{\mathbf{1},\mathbf{2}} \big)~ \mathcal{R} ~ \big(\mathcal{H}^\mu_{\mathbf{2},\mathbf{1}} \big)   }}{L^3~\sqrt{2E_\mathbf{1}}},
~\end{align}
where $\mathcal{R}$ is the finite-volume residue of the fully-dressed two-hadron propagator defined as 
\begin{equation}
\label{eq:Rintro}
 \mathcal{R}(E_{\mathbf{2}}, \textbf P) \equiv  \lim_{P_0 \to  E_{\mathbf{2}}} \left[ \frac{ ( P_0 - E_{\mathbf{2}}) }{F^{-1}(P,L) + \mathcal M(P)}\right] \,,
\end{equation}
where $F$ and $\mathcal M$ are the same objects appearing in the quantization condition above, Eq.~(\ref{eq:QC_master}). $\mathcal{R}$ is a matrix in the space of partial waves and open channels, and it can be constrained using the calculated finite-volume spectrum. Similarly, $\big(\mathcal{H}^\mu_{\mathbf{2},\mathbf{1}} \big)$ and $\big(\mathcal{H}^\mu_{\mathbf{1},\mathbf{2}} \big)$ are column and row vectors, respectively, in this same space.

This relationship exactly accounts, in a relativistic and model-independent way, for the strong interactions between hadrons in QCD up to corrections which scale like $\mathcal{O}(e^{-m_\pi L})$. The use of a single insertion of the vector current is accurate to first order of perturbation theory in QED.

\begin{figure}
\begin{center} 
\includegraphics[width=\columnwidth]{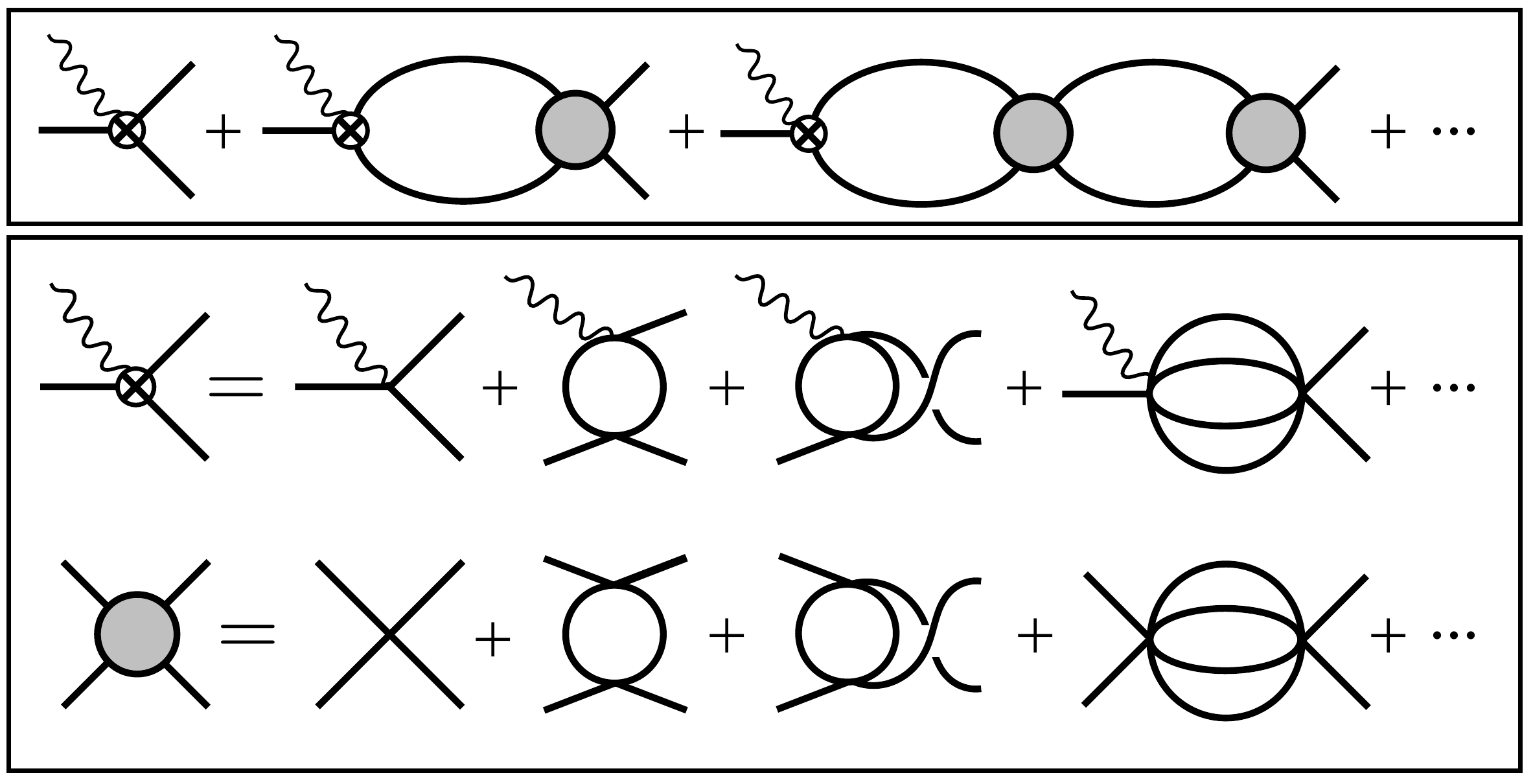}
\caption{The top line shows a diagrammatic representation~\cite{Briceno:2014uqa} of $\mathcal{A}_{\pi\pi, \pi\gamma^\star}$. Intermediate $\pi\pi$ propagators between the Bethe-Salpeter kernels (grey circles) are fully dressed, and the crossed circle is the fully-interacting vertex coupling $\pi$ to $\pi\pi$ in the presence of the external current. The vertex and the Bethe-Salpeter kernels are defined in the second and third lines, respectively. The rescattering series in the top line results in $\mathcal{A}_{\pi\pi, \pi\gamma^\star}$, which depends on the $\pi\pi$ scattering phase shift, and we see that, even for zero $\pi\pi$ rescattering, the amplitude need not be zero due to the initial production amplitude.   }
\label{fig:hmu}
\end{center}
\end{figure}

Similarly to the quantization condition, Eq.~\ref{eq:QC_master}, this relation reduces to a simple form when the lowest subduced partial wave is dominant. In Ref.~\cite{Dudek:2012xn} it was demonstrated that the $\pi\pi \to \pi \pi$ scattering amplitudes with $\ell\geq 3$ are negligibly small in the elastic scattering region. It does not necessarily follow from this that the transition amplitudes $\big(\mathcal{H}^\mu_{\pi \pi, \pi}\big)_{\ell \geq 3}$ are negligibly small -- as illustrated in Figure~\ref{fig:hmu}, there is a term due to the `production' amplitude which remains even in the case of no $\pi\pi$ rescattering. It can be argued though that we expect such production amplitudes for $\ell \geq 3$ to be kinematically suppressed at low-energy by a threshold barrier $\sim q^{\star \ell}$, and to be suppressed relative to the $\ell =1 $ amplitude which is dynamically enhanced by the resonant $\rho$. We will proceed assuming that only the $\ell=1$ transition plays a significant role -- see Appendix~\ref{sec:F_wave} for a discussion of the role a non-negligible $\ell=3$ amplitude might play.

Under the assumption of dominance of the $\ell = 1$ amplitude, we have
 \begin{align}
 \label{eq:Amp_to_matelem}
\big|{\mathcal{H}_{\pi\pi,\pi}^{\mu} }\big|
=  L^3 \sqrt{\frac{2E_\pi}{\mathcal{R}} } 
 \Big|{ \big\langle \pi,\Lambda_{\pi};L \big|{\mathcal{J}}^{\mu}(0) \big| \pi\pi,\Lambda_{\pi\pi};L \big\rangle} \Big|,
 \end{align}
where $\mathcal{R}$ is now a scalar given by
\begin{align}
\frac{2 E_\pi}{\RLL}
&= 32\pi \, \frac{E_\pi \Epipi}{q^{\star}_{\pi\pi}} \, \cos^2\delta_1 \nn\\
&\quad\quad\quad \times   \frac{\partial}{\partial P^\star_{0,\pi\pi}} {  \Big(\tan \delta_1+\tan\phi^{\mathbf{P}_{\!\pi \pi}, \Lambda_{\pi\pi}} \Big) }\bigg|_{ P^\star_{0,\pi\pi}=\Epipi^\star}\nn\\
&=
32\pi \frac{E_\pi \Epipi}{q^{\star}_{\pi\pi}}  \, \big(\delta_1'+r\phi'   \big),
\label{eq:LLfactorP}
\end{align}
where $\phi^{\textbf{P}_{\!\pi\pi}, \Lambda_{\pi\pi}}$ was defined in Eq.~(\ref{eq:Ppseudo-phase}) and
\begin{align}
r &\equiv   \cos^2\!\delta_1   \, / \,   \cos^2\!\phi^{\textbf{P}_{\!\pi\pi}, \Lambda_{\pi\pi}},   \nn \\
\delta_1'&\equiv
~\left.{\partial}{ \delta_1}/{\partial P^\star_{0,\pi\pi}} \right|_{ P^\star_{0,\pi\pi}=\Epipi^\star} , \nn \\
\phi'&\equiv
~\left.{\partial}{ \phi^{\textbf{P}_{\!\pi\pi}, \Lambda_{\pi\pi}} }/{\partial P^\star_{0,\pi\pi}} \right|_{ P^\star_{0,\pi\pi}=\Epipi^\star}.
\end{align}
The quantization condition, Eq.~(\ref{eq:QC}), implies that $r=1$, but we retain the form above when propagating statistical uncertainties on the spectrum energies though the calculation. These equations assume the hadrons in the ``$\mathbf{2}$'' state are distinguishable, as is appropriate for the process $\pi^+\pi^0\to \pi^+  \gamma^\star$ -- we discuss this further in Appendix~\ref{sec:symmetry_factors}.

These expressions, which depend only on the kinematics and dynamics of the $\pi\pi$ state, effectively leading to a proportionality between the finite and infinite-volume states, closely resemble the result for the $S$-wave derived by Lellouch and L\"uscher in their pioneering work, and as such we will refer to the inverse of $\RLL$ as the ``LL-factor''. As is evident, the LL-factor only depends on the nature of the finite-volume $\pi\pi$ state and is not particular to this production process. As a result, the LL-factor appearing here is the same as would appear in, for example, $\gamma^\star\to\pi\pi$~\cite{Meyer:2011um, Briceno:2015csa}. 
\footnote{We point the reader to Refs.~\cite{Feng:2014gba, Bulava:2015qjz} for recent numerical studies of this reaction.}

Since $\big( \mathcal{H}_{\pi\pi,\pi }^{\mu} \big)_{\ell = 1}$ has the Lorentz decomposition given in Eq.~\ref{eq:decomp}, using Eq.~\ref{eq:Amp_to_matelem} we can relate the finite-volume amplitude, $\tilde{\mathcal{A}}$, in Eq.~\ref{eq:mat_covariant}, to the infinite-volume amplitude, $\mathcal{A}_{\pi\pi, \pi \gamma^\star}$, by 
\begin{align}
\label{eq:AtoFtilde}
\big|{\mathcal{A}}_{\pi\pi,\pi\gamma^\star}(\Epipi^\star,Q^2) \big|=
\frac{\tilde{\mathcal{A}}(\Epipi^\star,Q^2; L)}{\sqrt{\RLL~2\Epipi}}.
\end{align} 

We could determine the infinite-volume amplitude using this relation directly, but it proves to be more convenient in this case, which features a narrow $\rho$ resonance and its corresponding rapid $E_{\pi\pi}^\star$ behavior, to proceed through an intermediate step where we write
\begin{equation}
\mathcal{A}_{\pi\pi, \pi \gamma^\star}(E_{\pi\pi}^\star, Q^2) = \frac{F(E_{\pi\pi}^\star, Q^2)}{\sqrt{2 E_{\pi\pi}^\star \, \mathcal{K}_{\pi\pi}(E_{\pi\pi}^\star) } } \, e^{i \delta_1(E_{\pi\pi}^\star)}.
\label{eq:AtoF}
\end{equation}
In this expression we have made an, at this stage, arbitrary division of the $E_{\pi\pi}^\star$ behavior into two real functions, $F(E_{\pi\pi}^\star, Q^2)$ and $\mathcal{K}_{\pi\pi}(E_{\pi\pi}^\star)$, and although only the magnitude appears in Eq.~\ref{eq:AtoFtilde} we have included for completeness the phase factor required to satisfy Watson's theorem.

\begin{figure}
\begin{center} 
\includegraphics[width=\columnwidth]{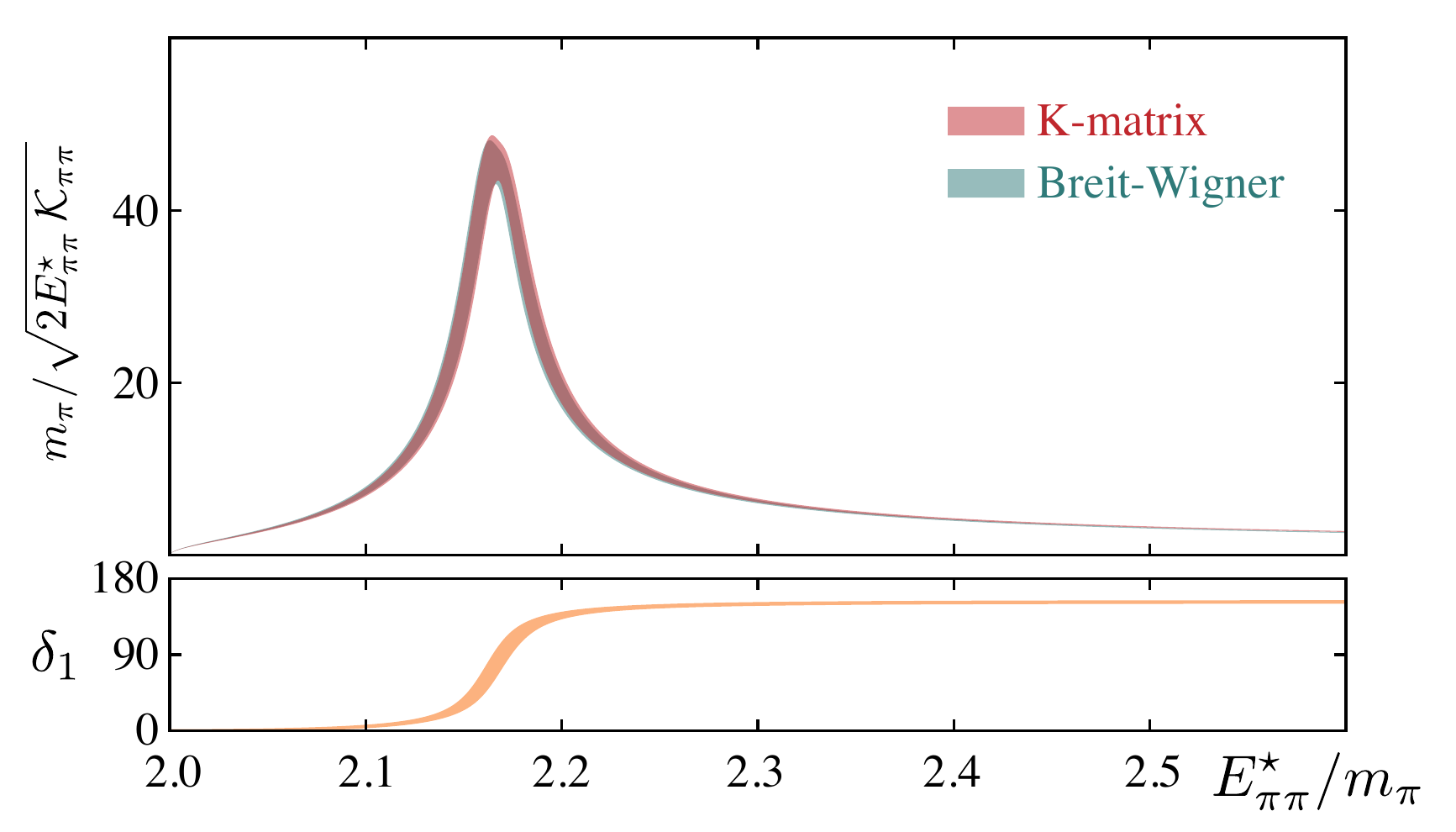}
\caption{Top panel shows $m_\pi/\sqrt{2 \Epipi^\star \Kpipi}$ as a function of the $\pi\pi$ energy, as defined in Eq.~(\ref{eq:AtoF}). The two parameterizations of the phase-shift given in Eqs.~\ref{eq:bw} and \ref{eq:kmat} are consistent and feature the expected enhancement of the transition amplitude in the vicinity of the $\rho$. Lower panel shows the $\pi\pi$ scattering phase shift for comparison.}
\label{fig:Kpipi}
\end{center}
\end{figure}

\begin{figure*}[t]
\begin{center}
\includegraphics[width=\textwidth]{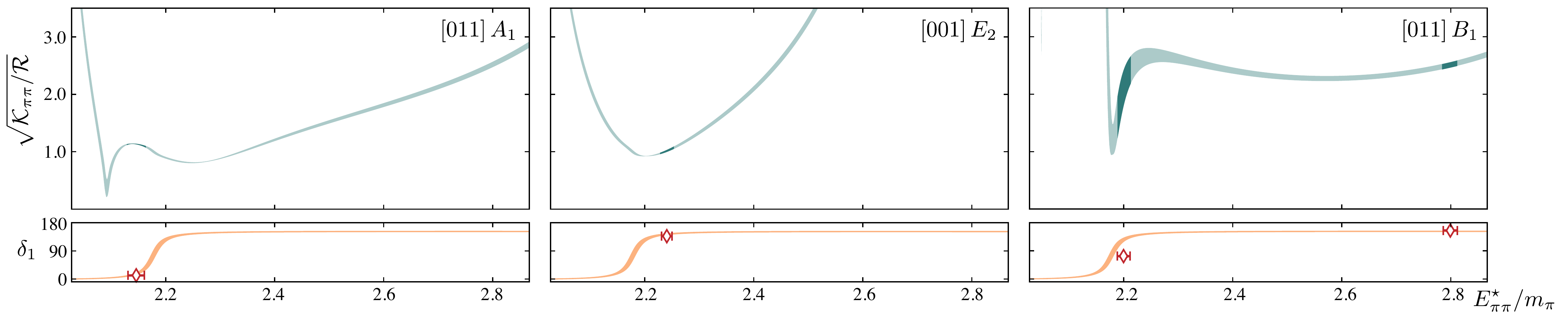}
\caption{Shown are examples of $\sqrt{\Kpipi/\mathcal{R}}$ for three $\pi\pi$ irreps. For the left and middle panels a Breit-Wigner parametrization, Eq.~(\ref{eq:bw}), of the scattering amplitude has been used, while for the right panel the $K$-matrix parametrization, Eq.~(\ref{eq:kmat}), has been used. The bands indicate the value of $\sqrt{\Kpipi/\mathcal{R}}$ as a function of the cm energy where the uncertainty is only due to that of the fit parameters in the phase-shift analysis. The darker regions indicate the position of the discrete finite-volume energies. Lower panels show the phase shift with the discrete values obtained for the corresponding irrep.} 
\label{fig:LL_factors}
\end{center}
\end{figure*}

We choose to parameterize $\mathcal{K}_{\pi\pi}(E_{\pi\pi}^\star)$ in a way which accounts for the sharply peaked resonance structure of the $\rho$, and in doing so we would expect $F(E_{\pi\pi}^\star, Q^2)$ to have only a modest residual $E_{\pi\pi}^\star$ dependence in the region of the $\rho$ resonance. We may write~\cite{Briceno:2015csa},
\begin{equation}
\label{eq:KF}
\frac{1}{\sqrt{ 2 E_{\pi\pi}^\star\; \mathcal{K}_{\pi\pi}(E_{\pi\pi}^\star) } } = \sin \delta_1(E_{\pi\pi}^\star) \, \sqrt{ \frac{16 \pi}{q^\star_{\pi\pi} \, \Gamma(E_{\pi\pi}^\star)}   },
\end{equation}
and we presented earlier two parameterizations, a Breit-Wigner form, Eq.~\ref{eq:bw} and a $K$-matrix form, Eq.~\ref{eq:kmat}, that can each describe the $P$-wave phase-shift in the elastic scattering region. It follows that 
\begin{equation}
F(E_{\pi\pi}^\star, Q^2) = \tilde{\mathcal{A}}(E_{\pi\pi}^\star, Q^2; L)\, \sqrt{ \frac{\mathcal{K}_{\pi\pi} }{\mathcal{R}} }, 
\end{equation}
and we find that while $\mathcal{K}_{\pi\pi}$ and $\mathcal{R}$ each change rapidly with $E_{\pi\pi}^\star$ in the $\rho$ resonance region, their ratio shows only modest dependence on $E_{\pi\pi}^\star$, and the strong correlation between their statistical fluctuations is reduced -- this is illustrated in Figures~\ref{fig:Kpipi} and \ref{fig:LL_factors}.

The decomposition in Eq.~\ref{eq:KF} is such that in the limit that $E_{\pi\pi}^\star$ approaches the $\rho$--pole, $F$ may be associated with the $\pi \rho$ transition form factor. Using Eq.~\ref{eq:KF}, we may rewrite Eq.~\ref{eq:AtoF} in a manner that makes this evident,
\begin{equation}
\mathcal{A}_{\pi\pi, \pi \gamma^\star}(E_{\pi\pi}^\star, Q^2) =\left( \frac{F(E_{\pi\pi}^\star, Q^2)}{\cot \delta_1(E_{\pi\pi}^\star)-i} \right)\,\sqrt{ \frac{16 \pi}{q^\star_{\pi\pi} \, \Gamma(E_{\pi\pi}^\star)}   }.
\label{eq:AtoF_pole}
\end{equation}
One observes that $\mathcal{A}_{\pi\pi, \pi \gamma^\star}$ has the same energy-dependent denominator as the elastic $\pi\pi$ scattering amplitude, and will have the same pole corresponding to the $\rho$. At the resonance pole, the residue of the  ${\pi\pi\to\pi\gamma^\star}$ amplitude factorizes into a product of couplings, $\pi\pi\to \rho$ and $\rho \to\pi\gamma^\star$, the latter in general being proportional to $F$ defined here. For larger quark masses, the $\rho$ becomes a stable hadron and the $\rho$-pole resides on the real $E_{\pi\pi}^\star$-axis below $\pi\pi$ threshold. In this limit the divergences in $\mathcal{R}$ and $\mathcal{K}_{\pi\pi}$ cancel exactly~\cite{Agadjanov:2014kha, Briceno:2015csa}. This is the scenario considered in, for example, Ref.~\cite{Shultz:2015pfa}. For quark-masses where the $\rho$ is unstable, the pole is complex and $F$ is still proportional to the residue of the $\pi\pi\to\pi\gamma^\star$ amplitude.

 Two of our $\pi\pi$ states, $(\mathbf{P}_{\!\pi\pi}=[011], B_1,n=0)$, and $(\mathbf{P}_{\!\pi\pi}=[111],E_2,n=0)$, are at energies where the phase-shift is very close to $90^\circ$, where $\mathcal{R}$ shows a large statistical uncertainty, leading to a disproportionately large uncertainty in $\frac{\mathcal{K}_{\pi\pi} }{\mathcal{R}}$ (see, for example, the third panel of Fig.~\ref{fig:LL_factors}). Given that this ratio must be equal to $1$ at the resonance mass, up to corrections of $\mathcal{O}(\Gamma_\rho/m_\rho)\sim \mathcal{O}(10^{-2})$~\cite{Briceno:2015csa}, we set $\frac{\mathcal{K}_{\pi\pi} }{\mathcal{R}} = 1$ here, while propagating uncertainties associated with the determination of the parameters appearing in Eqs.~\ref{eq:bw} and \ref{eq:kmat}. This is only a necessary approximation, applied for this pair of levels, because the $\rho$ is barely unstable at this quark mass. As the quark masses approach the physical point, the $\rho$ will become broader~\cite{Wilson:2015dqa} and this subtlety will disappear. For all other states we evaluate the LL-factor numerically and propagate its statistical and systematic uncertainties into the determination of the infinite-volume form factor and transition amplitude.

\begin{figure*}
\begin{center}
\includegraphics[width = \textwidth]{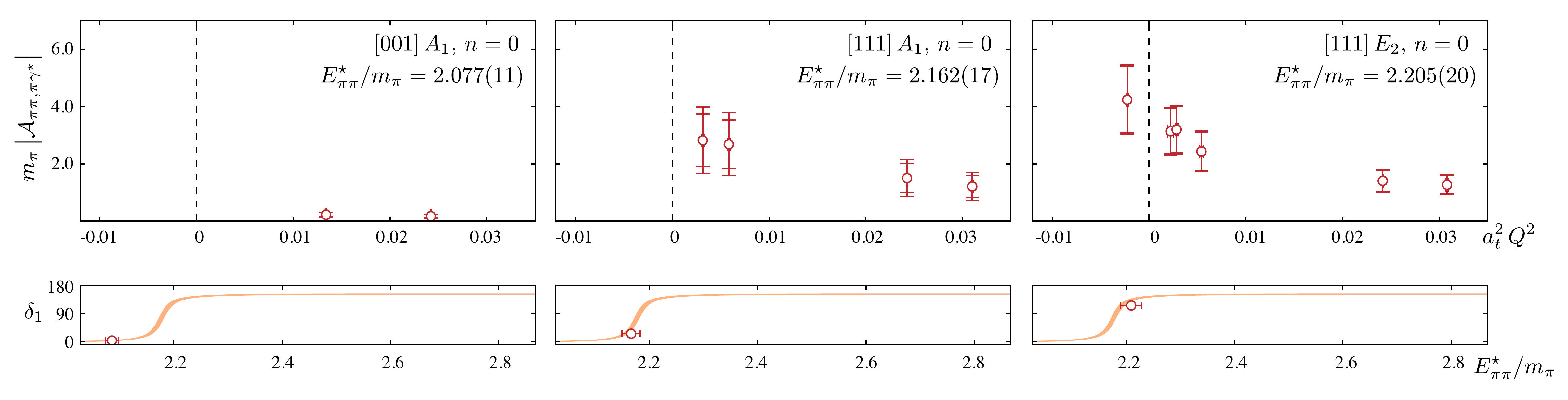}
\caption{Shown are three examples of the determined $\pi\pi\to\pi\gamma^\star$ transition amplitude, plotted in units of $m_\pi^{-1}$.  The momentum, irrep and eigenstate number $n$ are those of the $\pi\pi$ state. These three panels show the dynamical increase of the amplitude as $E_{\pi\pi}^\star$ moves through the resonant $\rho$. The inner and outer errorbars account for the statistical uncertainty on the three-point correlation functions, and the uncertainties in the $\pi\pi$ phase-shift parameterization parameters respectively. Lower panels show the phase shift with the discrete values obtained for the corresponding irrep. }
\label{fig:Apitopipi_irrep_plots}
\end{center}
\end{figure*}

\begin{figure*}[p]
\begin{center}
\includegraphics[width=\textwidth]{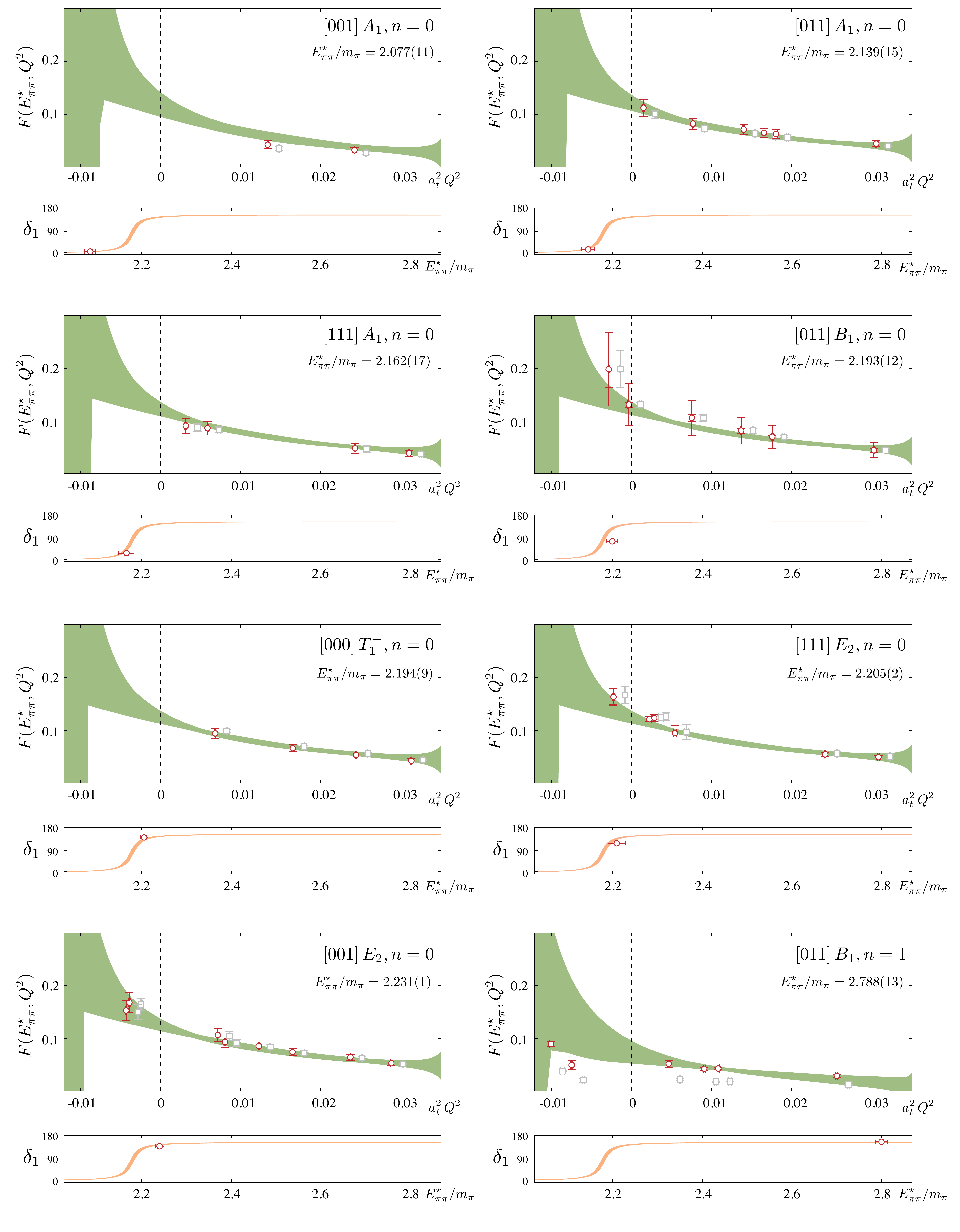}
 \caption{$F(E^\star_{\pi \pi}, Q^2)$ (red circles) for eight discrete $E_{\pi \pi}^\star$, extracted from $\tilde{A}(E^\star_{\pi \pi}, Q^2; L)$ using the $K$-matrix parametrization, Eq.~\ref{eq:kmat}, in $\sqrt{{\Kpipi}/{\RLL}}$ . $\tilde{A}(E_{\pi \pi}, Q^2; L)$ is also shown (grey squares, displaced in $Q^2$ for visibility) for comparison. The green band indicates the result of global fits to all $F(E^\star_{\pi \pi}, Q^2)$ values as described in the text.}
\label{fig:Fpirho_BW_irrep_plots}
\end{center}
\end{figure*}

\begin{figure}
\begin{center} 
\includegraphics[width=\columnwidth]{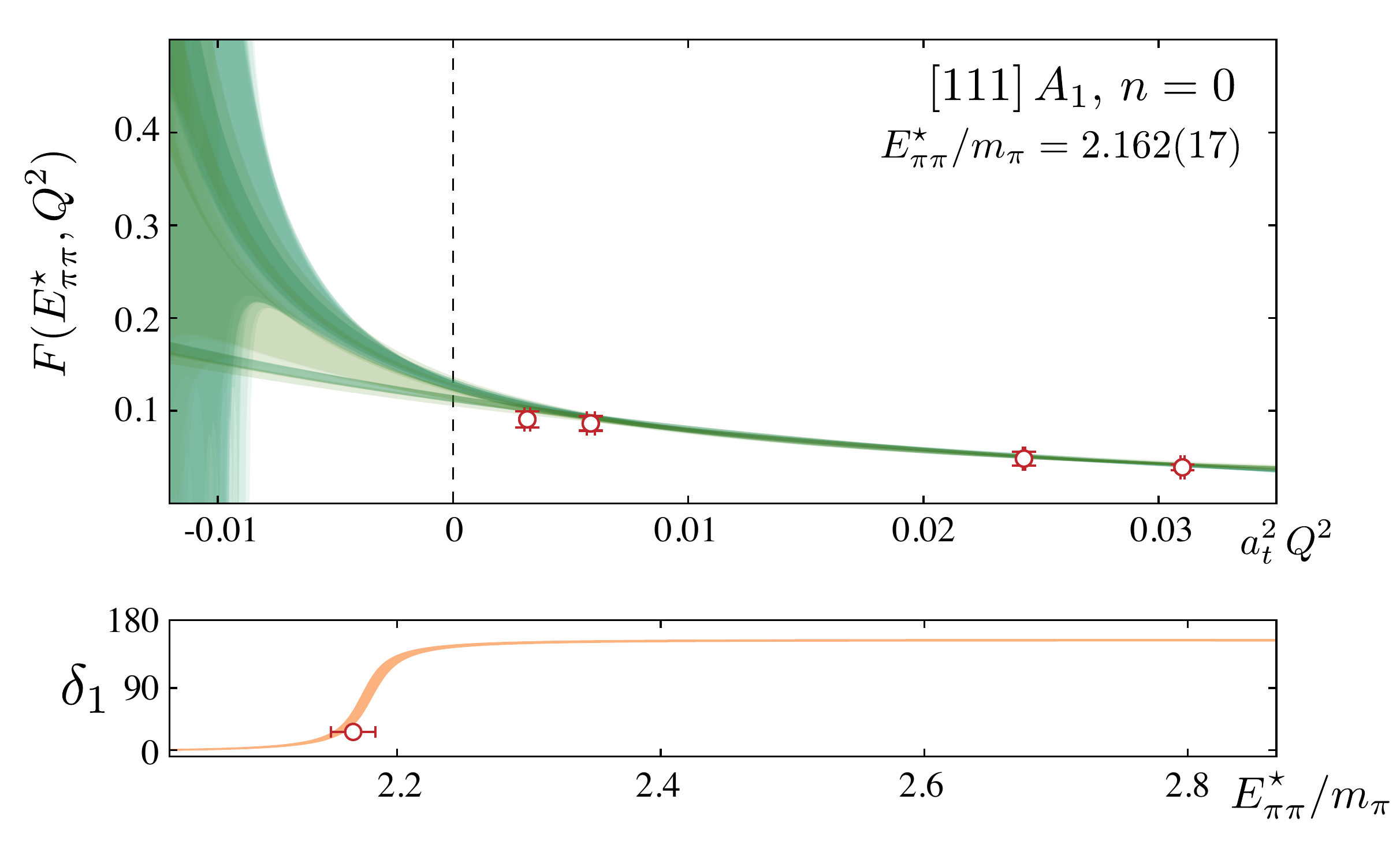}

\includegraphics[width=\columnwidth]{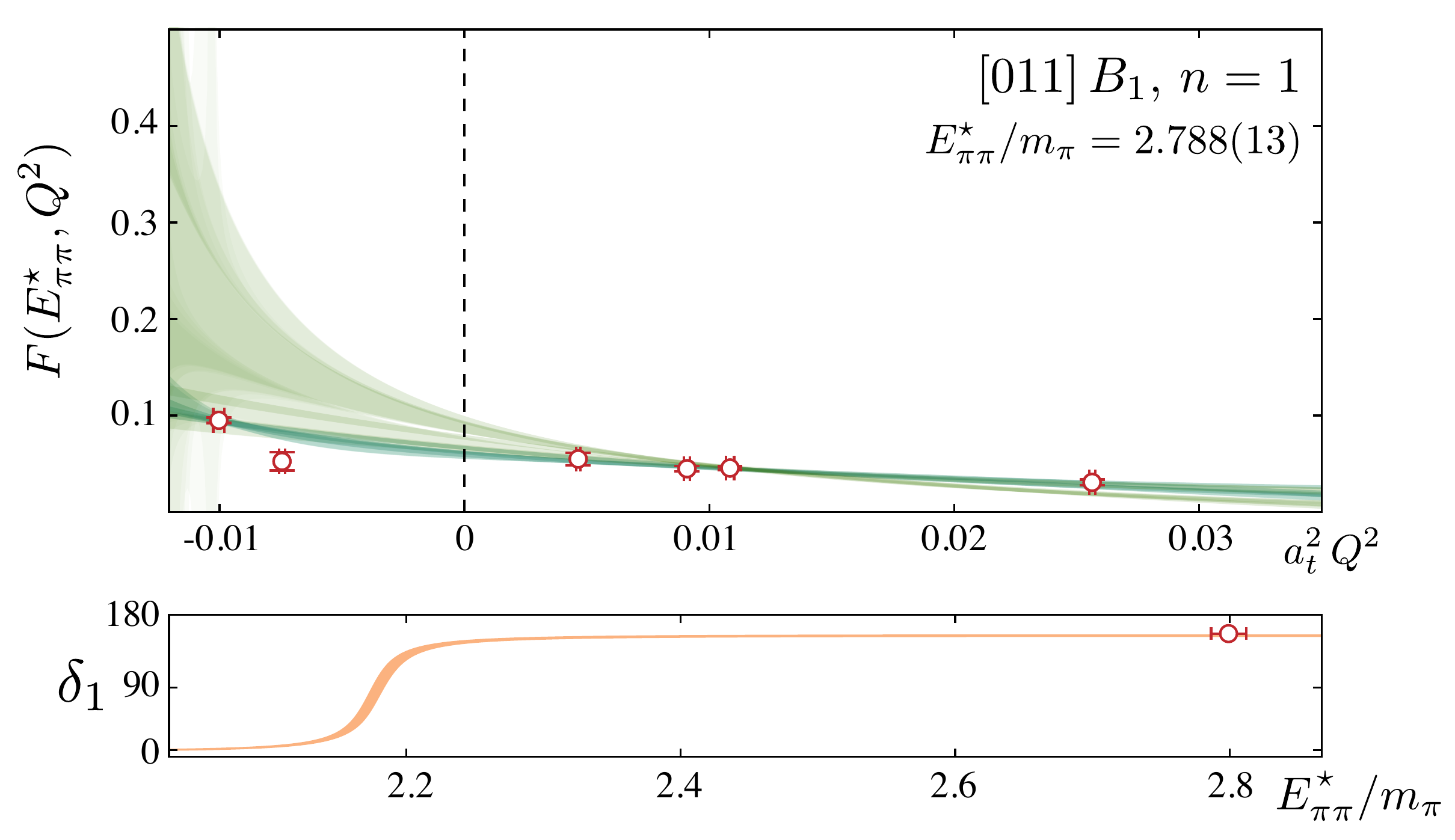}
\caption{Shown is a comparison of the fits of type {\sf A}, {\sf B} and {\sf C} (shades of green/blue), described in the text, that give a $\chi^2/N_\mathrm{dof} \leq 1.5$  for two representative example irreps.}
\label{fig:Fpirho_BW_band_plots}
\end{center}
\end{figure}

\section{Determination of the infinite volume transition amplitude}
\label{sec:global_fits}

With $\tilde{\mathcal{A}}(\Epipi^\star,Q^2,L)$ extracted from finite-volume three-point correlations functions and the Lellouch-L\"uscher factors evaluated using parameterizations of $\delta_1(E_{\pi \pi}^\star)$ which describe the finite-volume spectra, we may obtain the infinite volume $\pi\pi\to\pi\gamma^\star$ reduced amplitude, ${\mathcal{A}}_{\pi\pi,\pi\gamma^\star}$. In Fig.~\ref{fig:Apitopipi_irrep_plots} we give some examples of ${\mathcal{A}}_{\pi\pi,\pi\gamma^\star}$, plotted as a function of $Q^2$ for three values of $E_{\pi\pi}^\star$. We observe that this quantity has a strong dependence on $\Epipi^\star$ as expected, with significant increase in the transition amplitude observed at energies corresponding to the $\rho$ resonance. In the approach that we have taken, this resonant enhancement is present in the function $\mathcal{K}_{\pi\pi}(\Epipi^\star)$, with the $Q^2$ dependence residing in the form factor, $F(\Epipi^\star, Q^2)$, which shows only a mild dependence on $\Epipi^\star$. The form factor values, extracted when the $K$-matrix parameterization of $\delta_1$ is used, are presented in Figure~\ref{fig:Fpirho_BW_irrep_plots} -- the values extracted when the Breit-Wigner parameterization are equivalent within one standard deviation.

We can combine the kinematic points presented in Figure~\ref{fig:Fpirho_BW_irrep_plots} by performing a global fit of $F(\Epipi^\star, Q^2)$. 
We explore a flexible functional form,
\begin{align}
h^{[\{\alpha,\beta\}]}(\Epipi^\star,Q^2)&=\nn\\
&\hspace{-1cm}\frac{\alpha_1}{1+\alpha_2Q^2+\beta_1(\Epipi^{\star 2}-m^2_0)}+\alpha_3Q^2+\alpha_4Q^4
\nn\\
&\hspace{-1cm}+\alpha_5 \exp\left[{-\alpha_6 Q^2-\beta_2(\Epipi^{\star 2}-m_0^2) }\right]\nn\\
&\hspace{-1cm}+\beta_3(\Epipi^{\star 2}-m_0^2)+\beta_4(E^{\star4}_{\pi\pi}-m^4_0),
\label{eq:paramansatz}
\end{align}
where the $\alpha$'s and $\beta$'s are real-valued fit parameters, and $m_0$ is an arbitrary mass scale, which we set to $\mrho~m_\pi$ to coincide with real part of the $\rho$ resonance mass determined earlier. 

We consider a large number of fits in which we fix various $\alpha$'s and/or $\beta$'s to be zero. When all $\beta$'s are set to zero there is no $E_{\pi\pi}^\star$ behavior. The first term in Eq.~\ref{eq:paramansatz} allows for the possibility of a pole in $Q^2$ and the form is flexible enough to allow that pole's position to vary with $E_{\pi\pi}^\star$. We do not mean to imply any fundamental meaning to the form of this function, only that it is simple, flexible, and suitable to interpolate the data in $Q^2$ and $E_{\pi\pi}^\star$.

\begin{figure}[b]
\begin{center} 
\includegraphics[width=\columnwidth]{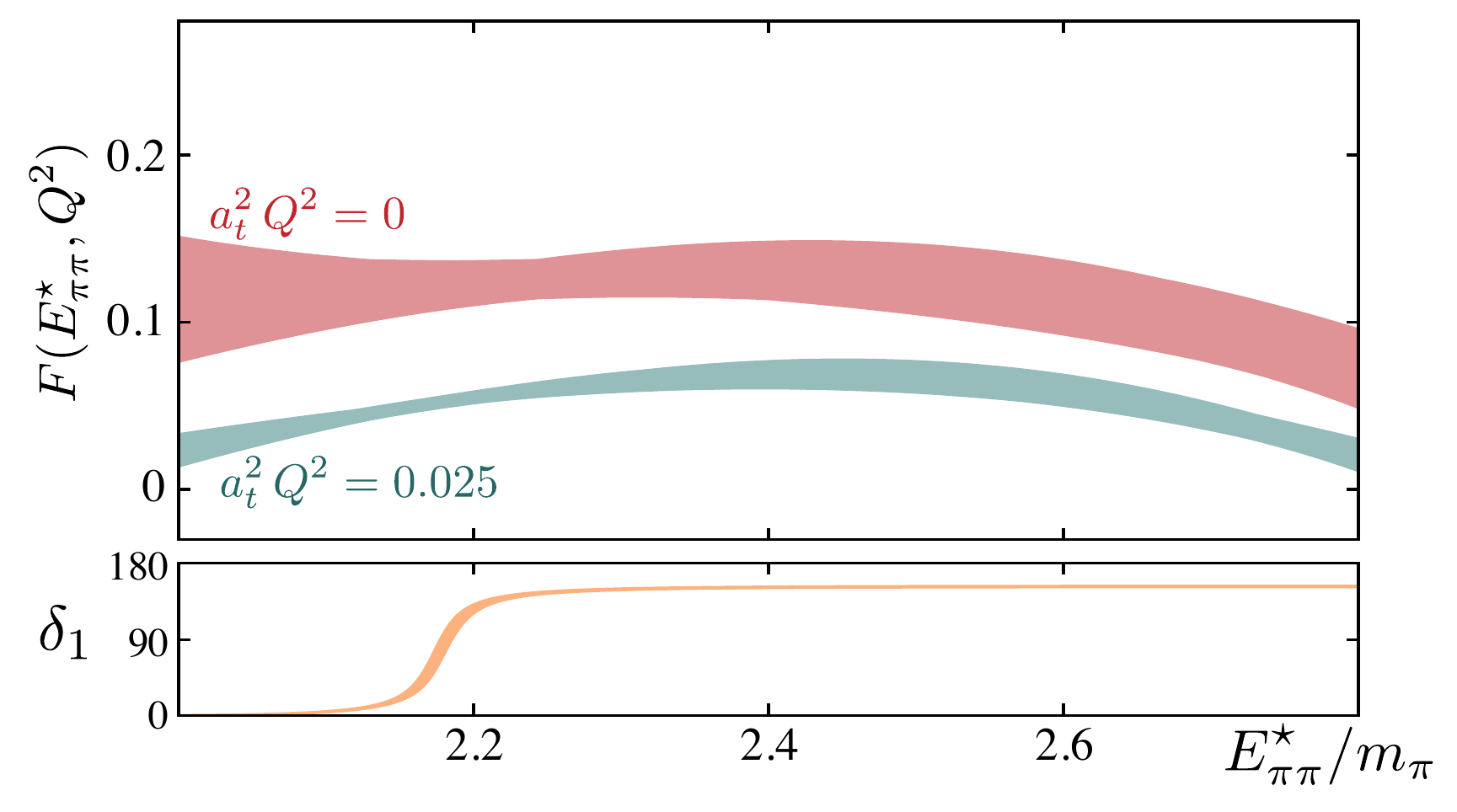} 
\caption{$F(E^\star_{\pi\pi}, Q^2)$ as a function of the $\pi\pi$ cm energy for two values of $a_t^2\, Q^2=0,0.025$. }
\label{fig:Fpirho_slices_in_Q2}
\end{center}
\end{figure}

\begin{figure}[t]
\begin{center} 
\includegraphics[width=\columnwidth]{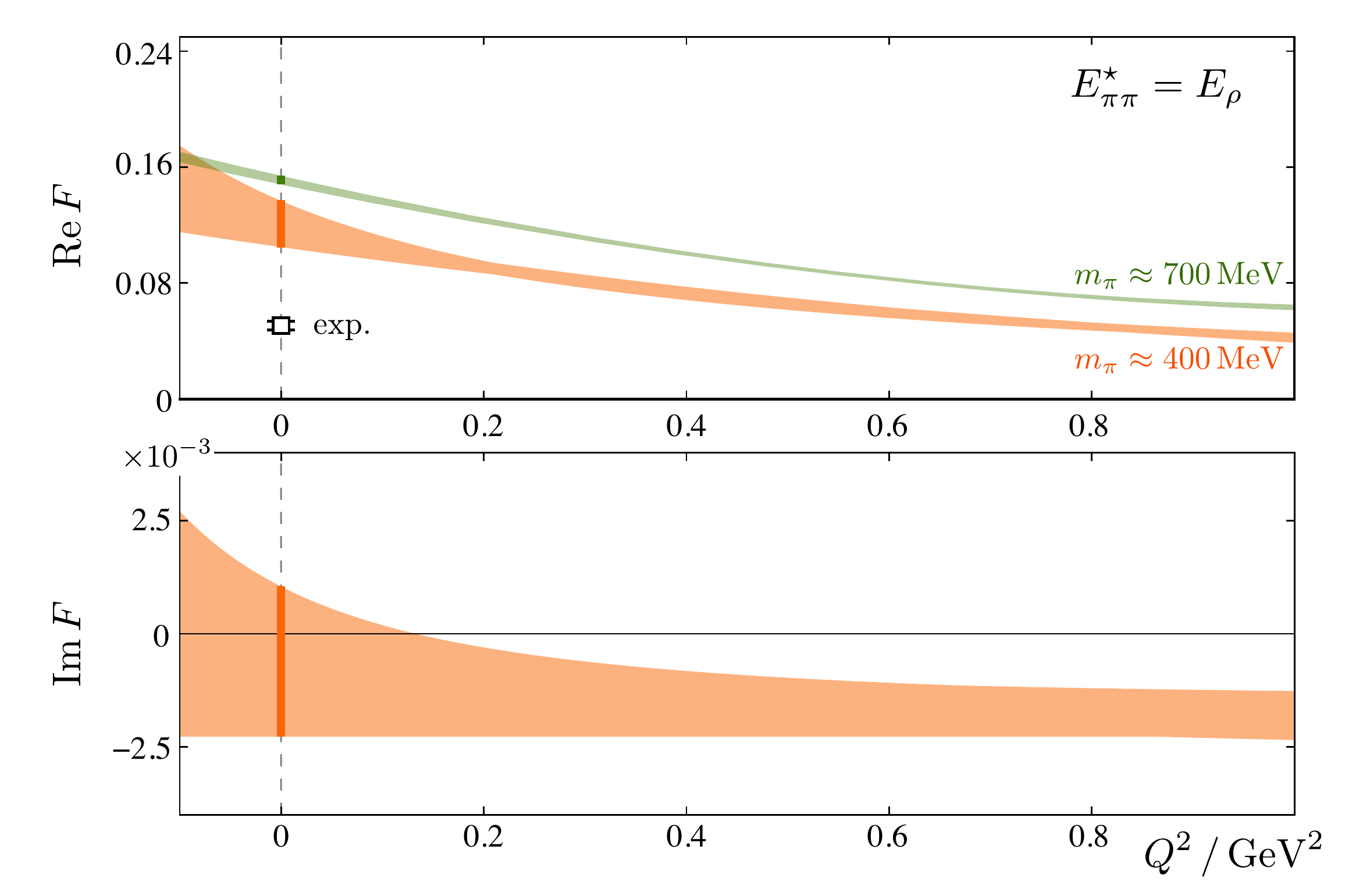}
\caption{The real and imaginary parts of the form factor determined in this work evaluated at the $\rho$ pole (orange). For comparison we show the form factor obtained in Ref.~\cite{Shultz:2015pfa} for a heavier quark mass, where the $\rho$ is stable (green). Also shown is the experimentally determined value for the $\rho\pi$ photocoupling~\cite{Huston:1986wi, Capraro:1987rp}.  }
\label{fig:Fpirho_Erho_BW}
\end{center}
\end{figure}

In performing fits, we define the data covariance matrix as $C_\mathrm{tot}=C_\mathrm{stat}+C_\mathrm{sys}$, where $C_\mathrm{stat}$ accounts for the statistical fluctuations over the ensemble of configurations in this calculation, while $C_\mathrm{sys}$ accounts for the uncertainty in the fit parameters used to describe $\delta_1(E_{\pi\pi}^\star)$.

\begin{figure*}
\begin{center} 
\includegraphics[width=0.7\textwidth]{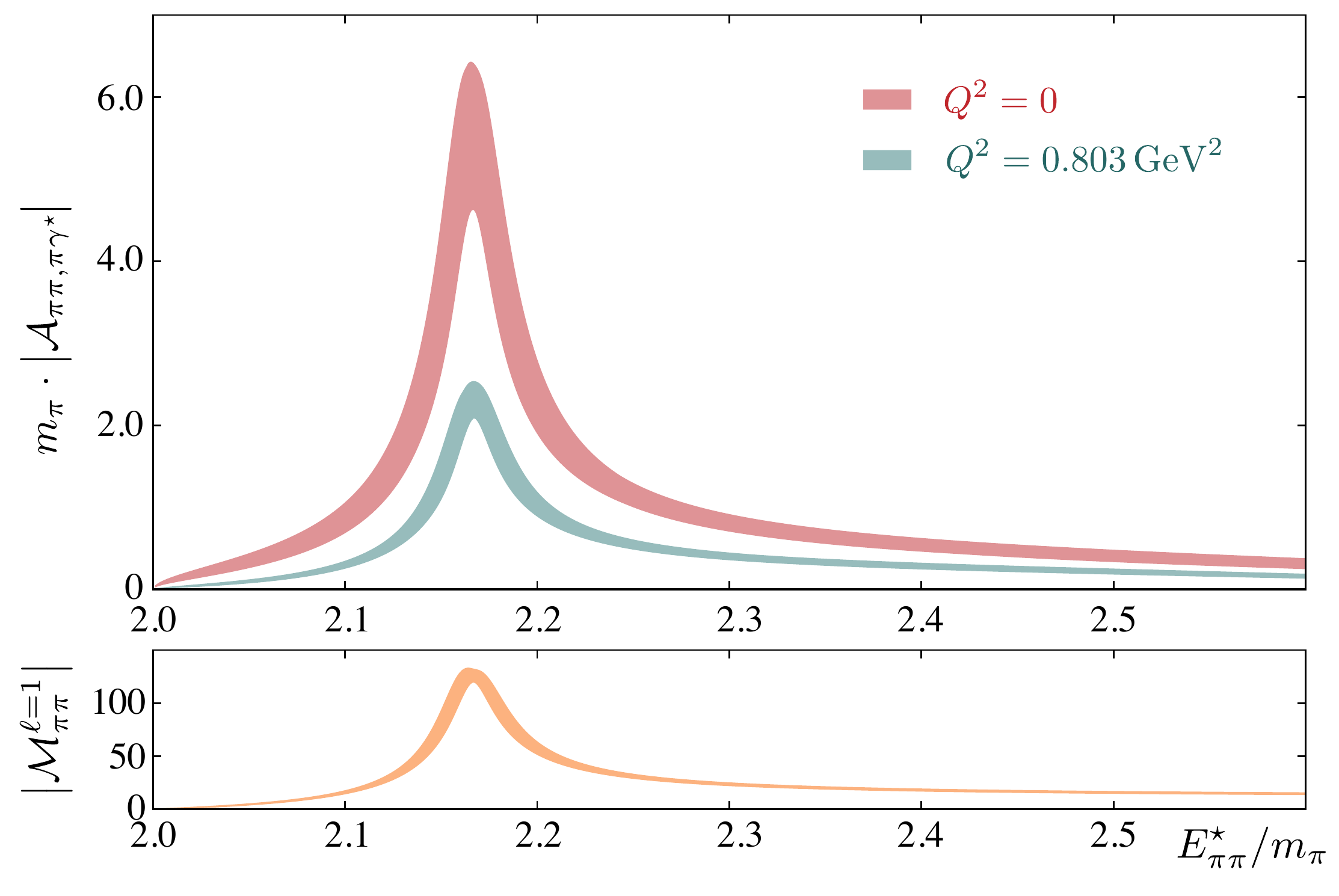}
\caption{$m_\pi |\Amp|$ for two values of $Q^2$ as a function of $\Epipi^\star$ along with the elastic $\pi\pi$ $P$-wave scattering amplitude. }
\label{fig:slices_in_Q2}
\end{center}
\end{figure*}

\begin{figure}
\begin{center}
\includegraphics[width=\columnwidth]{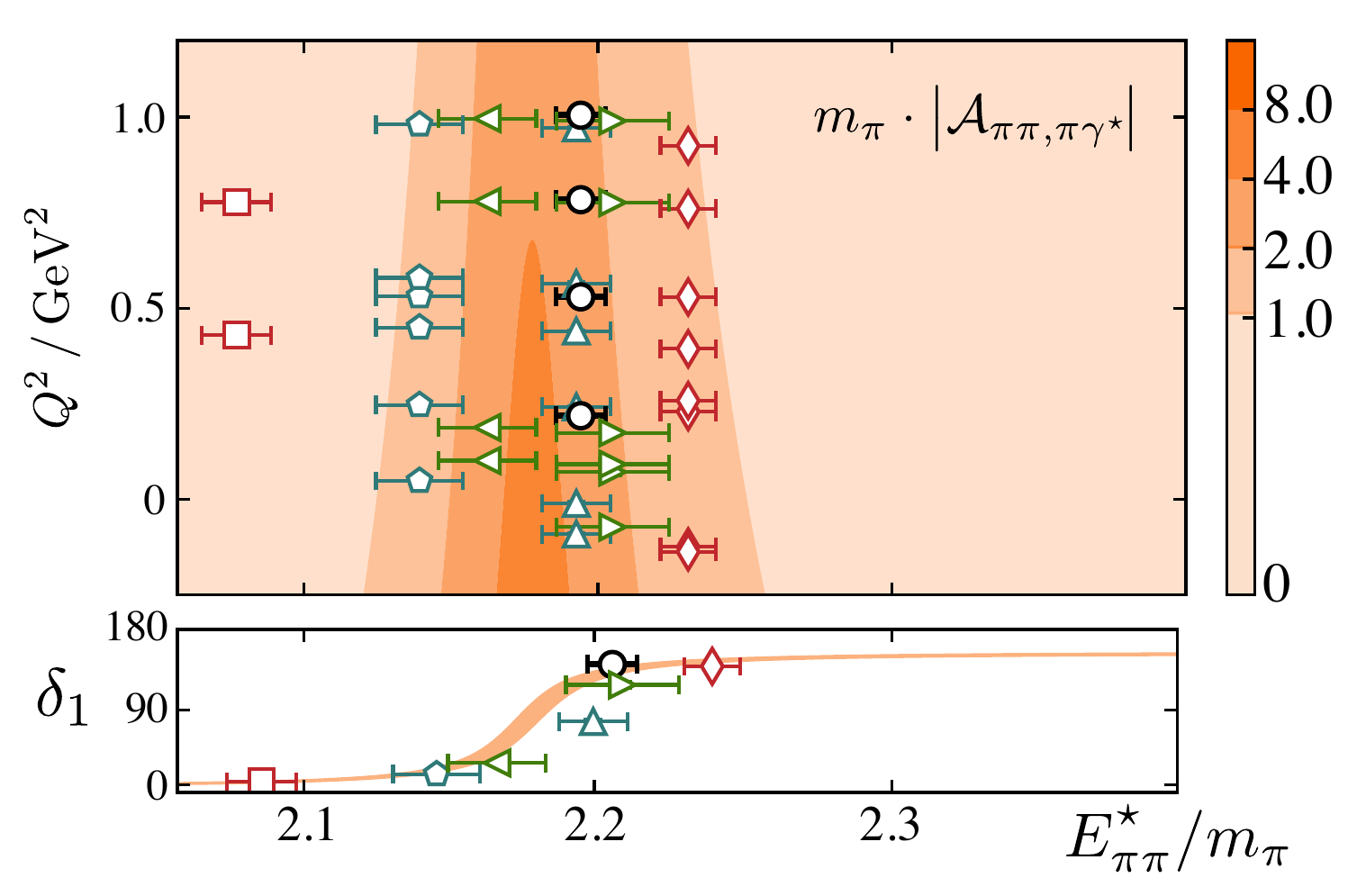}
\caption{Mean value of $m_\pi |\Amp|$ plotted as contours along with the locations of the points $(\Epipi^\star/m_\pi,Q^2)$ where the finite-volume matrix elements were determined. A total of 42 different kinematic points were used, and 6 of these appear outside the range plotted here.}
\label{fig:Apipi_to_pi_abs}
\end{center}
\end{figure}

\begin{figure*}
\begin{center}
\hspace*{-.7cm}                                                           
\includegraphics[width=0.7 \textwidth]{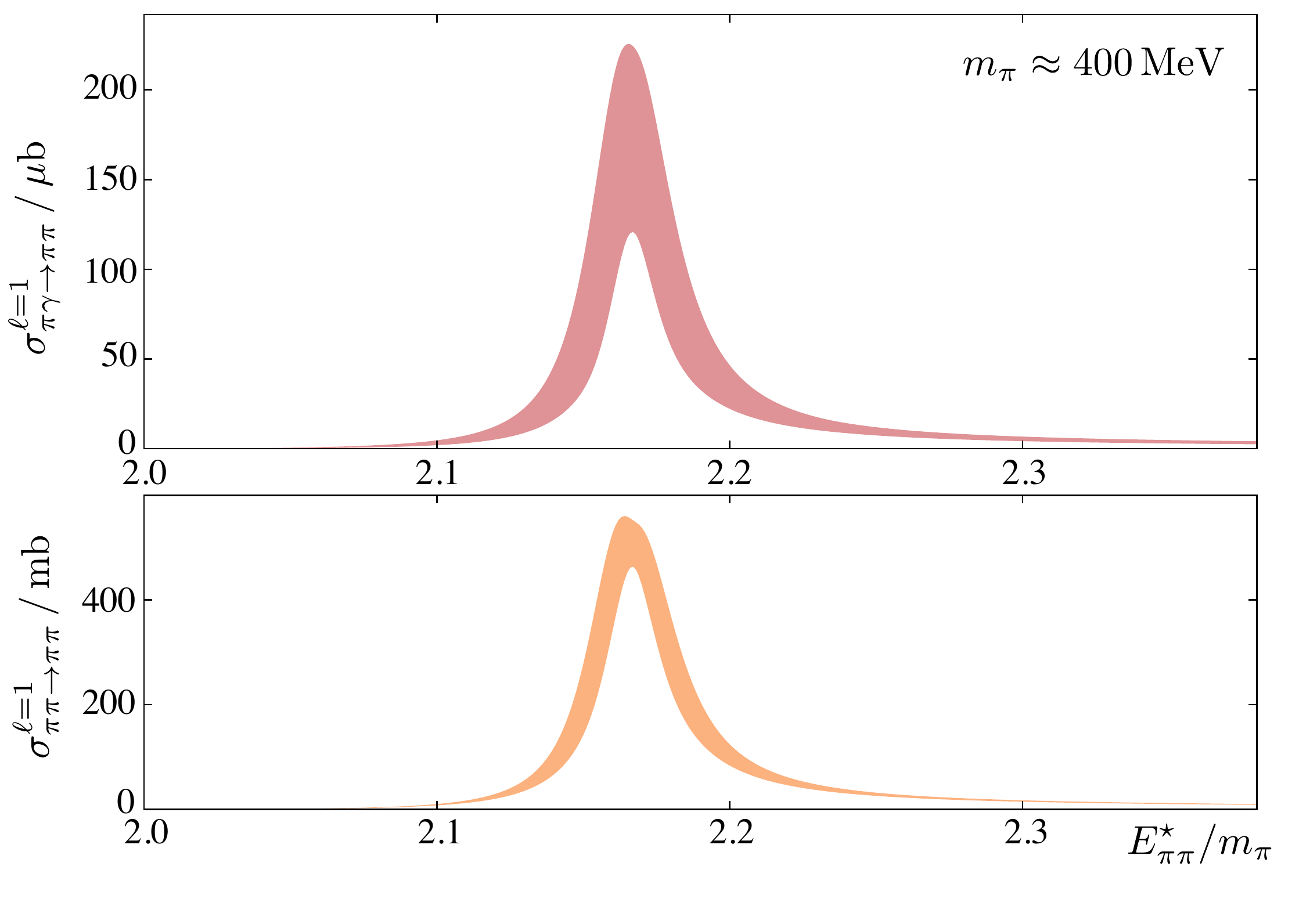}
\caption{$\pi^+\gamma\to\pi^+\pi^0$ cross section as a function of the $\pi\pi$ cm energy along with the $\ell=1$ elastic $\pi\pi$ scattering cross section. The $\rho$ resonance is clearly visible in both cross-sections.}\label{fig:cross_sec}
\label{fig:cross_section}
\end{center}
\vspace*{-.5cm}
\end{figure*}

The green bands in Figure~\ref{fig:Fpirho_BW_irrep_plots} show the result of global fits, restricting to fits that provide a good description of the data ($\chi^2/\mathrm{dof}\leq1.5$). All successful fits are found to require some $E_{\pi\pi}^\star$ dependence in $F( E_{\pi\pi}^\star, Q^2)$. In Figure~\ref{fig:Fpirho_BW_band_plots} we present examples of the results of three different types of fits: types {\sf A}, {\sf B} and {\sf C}.  Type {\sf A} fits correspond to using the full set of data points,  and restricting the $Q^2$ pole in Eq.~(\ref{eq:paramansatz}) to be independent of $E_{\pi\pi}^\star$ ($\beta_1=0$). Type {\sf B} fits include all data points and do allow for a pole in $Q^2$ to depend on $E_{\pi\pi}^\star$. Type {\sf C} fits are those in which we prune the data set by excluding time-like ($Q^2 < 0$) points. We conclude that we have insufficient time-like data points to strongly constrain the position of any possible pole in $Q^2$. The green bands in Figure~\ref{fig:Fpirho_BW_irrep_plots} conservatively encompass the range of behaviors given by all successful fits of type {\sf A}, {\sf B} and {\sf C}, and it is clear that this assessment leads to only a moderate overall uncertainty in the space-like region where the form factor is rather well constrained.

This procedure is repeated for the Breit-Wigner parametrization of the phase shift leading to very similar results -- in what follows we account for the small difference between the two parameterizations in our systematic uncertainty.

Figure~\ref{fig:Fpirho_slices_in_Q2} illustrates, for two values of $Q^2$, the mild $E_{\pi\pi}^\star$ behavior found for $F(E_{\pi\pi}^\star, Q^2$). This suggests that this function has a very mild dependence on $E_{\pi\pi}^\star$ for a large kinematic region. We have determined this function for seven energies below $E_{\pi\pi}^\star= 2.3~m_\pi$ and another one at $E_{\pi\pi}^\star= 2.8 ~m_\pi$. Therefore, it is possible that our interpolation does not reliably describe this function in the region between $E_{\pi\pi}^\star= 2.3~m_\pi$ and $2.8 ~m_\pi$, although there is no reason to expect stronger energy-dependence.

A rigorous way to define the electromagnetic transition form factor for $\rho \to \pi \gamma^\star$ is to take the amplitude $\mathcal{A}_{\pi\pi, \pi \gamma^\star}(E_{\pi\pi}^\star, Q^2)$, constrained at real values of $E_{\pi\pi}^\star$, and analytically continue it to the pole in the complex plane at $E_{\pi\pi}^\star = E_\rho = \big(\Erho \big) \, m_\pi$. As made evident by Eq.~\ref{eq:AtoF_pole}, the residue of $\mathcal{A}_{\pi\pi, \pi \gamma^\star}$ at the pole can be factorized into a product of couplings of the $\rho$ to $\pi\pi$ and to $\pi\gamma^\star$ where the second of these will be proportional to $F(E_\rho, Q^2)$. In Figure~\ref{fig:Fpirho_Erho_BW} we show this quantity, where the orange band encompasses all satisfactory fits described previously using both parameterizations of the $\pi\pi$ phase-shift. The smallness of the imaginary part is due to the $\rho$ pole at this quark mass being rather close to the real energy axis and the energy dependence of $F(E_{\pi\pi}^\star, Q^2)$ being rather mild. Figure~\ref{fig:Fpirho_Erho_BW} also shows (in green) the form factor of the $\rho$ computed with a heavier light quark mass such that the pion has mass $\sim 700\,\mathrm{MeV}$ and the $\rho$ is a stable hadron~\cite{Shultz:2015pfa}. We also compare to experimental estimates of the real part of the $\rho\pi$ photocoupling~\cite{Huston:1986wi, Capraro:1987rp}. In Eq.~\ref{eq:decay_rate} we give the relation between this definition of the form factor and the radiative decay width of $\rho^+\to\pi^+\gamma$.

In performing the analytic continuation of $\mathcal{A}_{\pi\pi, \pi \gamma^\star}$ as a function of $E_{\pi\pi}^\star$, we have kept the masses of all external hadrons fixed at their on-shell values. Furthermore, we have explored only real virtualities for the photon. Such an approach mirrors existing determinations~\cite{Workman:2013rca,Ronchen:2014cna,Svarc:2014sqa} of pion photoproduction residues from experimental measurements of $N \gamma \to N^\star \to \pi N$. We believe this is a natural choice for general virtualities if one identifies $Q^2=-m^2_\gamma$. An alternative extrapolation procedure was presented in Ref.~\cite{Agadjanov:2014kha}, where the authors suggest determining $\mathcal{A}_{\pi\pi, \pi \gamma^\star}$ for a range of values of $E_{\pi\pi}^\star$ while fixing $|\textbf{Q}|$ in the c.m. frame of the $\pi \gamma^\star$ state. One can then extrapolate $E_{\pi\pi}^\star$ to the $\rho$ pole while keeping $|\textbf{Q}|$ fixed. The advantage of this procedure is that one does not need to perform a global fit of the amplitude in terms of the variable $Q^2$. However this procedure has not been described for the most useful means of accessing a large number of energy levels in a finite volume, utilized in this paper, namely boosting of the $\pi\pi$ system to non-zero total momentum.


With a determination of $F(E^\star_{\pi\pi},Q^2)$ in hand we may construct the $P$-wave reduced amplitude, ${\mathcal{A}}_{\pi\pi,\pi\gamma^\star}(E^\star_{\pi\pi},Q^2)$, using Eqs.~(\ref{eq:AtoF}, \ref{eq:KF}). Since the phase of the amplitude is fixed by Watson's theorem to match the $\pi\pi$ phase, we only report its magnitude, which we choose to present in units of $m_\pi^{-1}$. In Fig.~\ref{fig:slices_in_Q2}, we show the result for the transition amplitude as a function of $E_{\pi\pi}^\star$ for two values of $Q^2$ along with the elastic $\pi\pi$ scattering amplitude, $\mathcal{M}_{\pi\pi}^{\ell=1}$. The bands shown encompass all the 1$\sigma$ fluctuations obtained using various different parameterizations and hence can be considered to include both statistical and systematic error estimates. Figure~\ref{fig:Apipi_to_pi_abs} makes clear that our determination of the amplitude has been constrained by points which sample well the entire relevant region of $E_{\pi\pi}^\star$ and $Q^2$.

\section{$\pi^+\gamma\to\pi^+\pi^0$ cross section}
\label{sec:cross_section}

Having obtained the transition amplitude, we can proceed to determine the dominant $P$-wave contribution to the $\pi^+\gamma\to\pi^+\pi^0$ cross section, which can readily be compared with phenomenological studies~\cite{Kaiser:2008ss, Hoferichter:2012pm}. For simplicity, we restrict our attention to the process where the incoming photon is on-shell, $Q^2=0$, but all results generalize to describe the dominant one-photon exchange contribution to the $\pi^+e^-\to\pi^+\pi^0e^-$ cross section.

In Appendix~\ref{app:cross_section} we show that,
\begin{align}
\sigma^{\ell=1}(\pi^+\gamma\to\pi^+\pi^0) 
&= 
\alpha
\frac{{q}^\star_{f} \, {q}^\star_{i}}{m_{\pi}^2}  
\big| {\mathcal{A}}_{\pi\pi,\pi\gamma^\star}(\Epipi^{\star},0) \big|^2,
 \label{eq:cross_section}
\end{align}
where ${q}^\star_{i}$ and ${q}^\star_{f}$ are the cm-frame momenta in the initial and final states, respectively. Similarly, the $\pi\pi$ elastic scattering cross-section due to the $P$-wave is given by
\begin{align}
\label{eq:cs_and_ps}
\sigma^{\ell=1}(\pi^+\pi^0\to\pi^+\pi^0)=\frac{12\pi}{q^{\star2}}~\sin^2\delta_1,
\end{align}
where $q^\star$ is the cm-frame momentum. 

In Fig.~\ref{fig:cross_section} we plot both cross-sections for comparison. We observe both the elastic scattering and the radiative transition cross sections are dynamically enhanced in the same region of energy due to the presence of the $\rho$ resonance, and we see the reduction in magnitude expected for the electromagnetic process relative to the strong process. 

Comparing $\sigma^{\ell = 1}(\pi^+\gamma\to\pi^+\pi^0)$ to the phenomenological cross-section~\cite{Kaiser:2008ss, Hoferichter:2012pm}, we find that the peak cross-section in our calculation with $m_\pi\approx 400$~MeV is nearly one order of magnitude larger than those in Refs.~\cite{Kaiser:2008ss, Hoferichter:2012pm}. This apparent discrepancy can be understood by investigating the dependence of the peak cross section on the width of the resonance (see Eq.~(\ref{eq:cross_section_param2})),
\begin{align}
\lim_{E_{\pi\pi}^\star\to m_\rho}
\sigma^{\ell=1}(\pi^+\gamma\to\pi^+\pi^0) \propto  \frac{q^\star_i \F^2(m_{\rho},0)}{m_{\pi}^2~\Gamma_P(m_{\rho})}.
\label{eq:cross_section_peak}
\end{align}
From Fig.~\ref{fig:Fpirho_Erho_BW}, we find that $q^\star_i\F^2(m_{\rho},0)/m_{\pi}^2$ is approximately 60\% of the experimental value. With the two quark-mass points at our disposal, we can speculate that the quark-mass dependence of this quantity is relatively mild. Meanwhile, the $\rho$ width is known to depend strongly on the quark mass and for the quark masses used here it is around 12 MeV~\cite{Dudek:2012xn}, making it an order of magnitude smaller than experiment~\cite{pdg:2014}. It reasonable to expect that for calculations performed with decreasing values of the quark masses, the $\rho$-resonance will become broader (see Ref. \cite{Wilson:2015dqa} for a concrete example at $m_\pi \approx 230$~MeV), and the $\pi^+\gamma\to\pi^+\pi^0$ cross section will decrease significantly.

 \section{Conclusion  and outlook \label{sec:conclusion}}
In this paper we have described the first calculation of the radiative decay of a resonance within a first-principles approach to QCD. By computing three-point correlation functions using lattice QCD we determine $\pi\pi\to\pi\gamma^\star$ matrix elements in a finite-volume over a range of discrete kinematic points. These are related to the corresponding infinite-volume transition amplitude using a procedure which features the $\pi\pi$ elastic scattering amplitude determined from the discrete spectrum of states on the same lattice configurations. The $P$-wave amplitude for $\pi\pi\to\pi\gamma^\star$ is found to feature a dynamical enhancement corresponding to the $\rho$ resonance, and the residue of the amplitude at the $\rho$-pole can be used to determine the $\rho\to\pi\gamma^\star$ transition form factor.

In the present calculation we made a small number of approximations which will be addressed in subsequent studies. We used only a single lattice volume, but the formalism should give compatible results for any volume large enough that exponentially suppressed corrections of the form $e^{-m_\pi L}$ can be neglected. For the $\pi\pi$ spectrum these corrections have been studied analytically~\cite{Albaladejo:2013bra, Bedaque:2006yi} and demonstrated to be small, but they have not been explored for transition amplitudes. Future calculations using multiple volumes will address this.

A recent determination of the $P$-wave $\pi\pi$ elastic scattering amplitude at a lighter pion mass, ${m_\pi \approx 230 \, \mathrm{MeV}}$~\cite{Wilson:2015dqa}, shows the expected decrease in $\rho$ mass and increase in decay width, and an application of the methods outlined in this paper to the same ensemble of lattice configurations is now warranted.

A possible step once the transition amplitudes are evaluated at a few quark masses is to consider a chiral extrapolation of these quantities, in order to make more direct contact with experimental observables, in advance of an eventual calculation at the physical pion mass. Currently, it is not completely clear how such an extrapolation could be performed. The necessary formalism that accommodates resonances and that incorporates quark-mass dependence in a transition process featuring an external current is missing, unlike the case of elastic and inelastic meson-meson scattering amplitudes~\cite{Oller:1997ng, Dobado:1996ps, Oller:1998hw, GomezNicola:2001as, Pelaez:2006nj} (recently implemented in the analysis of $\pi\pi$ elastic scattering~\cite{Bolton:2015psa}). One possible method which potentially may reduce the systematic uncertainty associated with describing the $(E_\pi\pi^\star, Q^2)$ dependence of the amplitude and could allow a constrained chiral extrapolation, is to make use of amplitudes obtained using dispersive techniques~\cite{Hoferichter:2012pm}.

Beyond being a physically interesting process in its own right, $\pi\pi \to \pi \gamma^\star$ serves as the first example of a wide class of phenomenologically important processes that can be studied with the techniques applied for the first time in this paper. The calculation presented here makes it clear that matrix elements featuring resonating hadronic systems can be rigorously studied using lattice QCD. Obvious extensions include nucleon resonances like the $\Delta$ in $\gamma^\star N \to \Delta\to N\pi$~\cite{Alexandrou:2007dt, Alexandrou:2007xj, Alexandrou:2010uk}, and heavy flavor decays which feature resonances, like $B\to \pi\pi \ell\nu$~\cite{Bowler:2004zb}. Moving to higher mass resonances, the extension into the coupled-channel case, accommodated by the formalism laid down in Refs.~\cite{Briceno:2015csa, Briceno:2014uqa, Briceno:2015tza}, will eventually allow calculations of radiative transitions featuring the exotic hybrid mesons that it is hoped will be photoproduced in the GlueX experiment~\cite{Dudek:2012vr, Ghoul:2015ifw}.

\noindent
\subsection*{Acknowledgments}

We thank our colleagues within the Hadron Spectrum Collaboration. 
The software codes
{\tt Chroma}~\cite{Edwards:2004sx} and {\tt QUDA}~\cite{Clark:2009wm,Babich:2010mu} were used to perform this work on clusters at Jefferson Laboratory under the USQCD Initiative and the LQCD ARRA project. We acknowledge resources used at the Oak Ridge Leadership Computing Facility, the National Center for Supercomputing Applications, the Texas Advanced Computer Center and the Pittsburgh Supercomputer Center. 
R.B., J.J.D. and R.G.E. acknowledge support from the U.S. Department of Energy contract DE-AC05-06OR23177, under which Jefferson Science Associates, LLC, manages and operates the Jefferson Lab. J.J.D. acknowledges support from the U.S. Department of Energy Early Career award contract DE-SC0006765. C.E.T. was partially supported by the U.K. Science and Technology Facilities Council [grant number ST/L000385/1]. C.E.T. and D.J.W. acknowledge support from the Isaac Newton Trust/University of Cambridge Early Career Support Scheme [RG74916]. R.B. would like to thank M. Hansen, A. Rusetsky, S. Sharpe, W. Detmold,  I. V. Danilkin, Z. Davoudi, J. Goity, M. Pennington, and S. Meinel for useful discussions.

\noindent

\appendix


\section{Notational conventions and normalizations\label{sec:convention}}
Quantities associated with a given channel carry a subscript labelling the channel, for instance the four-momentum of the $``\pi\pi"$ state is $P_{\pi\pi}$. Similarly, the total energy of the $``\pi"$ state is $E_\pi$. Quantities evaluated in the cm-frame carry a superscript star, e.g. $E_{\pi\pi}^\star$.

While infinite-volume single-hadron states with continuous three-momentum are normalized using the standard relativistic prescription, namely, 
\begin{align}
\label{eq:infty_states}
\big\langle  \pi,P_\pi \big| \pi, P_\pi' \big\rangle=2E_{\pi} \, \delta^3(\textbf P_{\!\pi}-\textbf P'_{\!\pi}),
\end{align} 
finite-volume states with discrete three-momentum are normalized to unity, 
\begin{align}
\big\langle  \pi,P_\pi;L \big| \pi, P_\pi';L \big\rangle=\delta_{\textbf P^{}_{\!\pi},\textbf P'_{\!\pi}}.
\end{align} 
 
The expansion in partial-waves of infinite-volume two-pion states follows that of Refs.~\cite{Briceno:2015csa, Briceno:2014uqa},
 \begin{equation}
\big| P_{\pi\pi},\hat{\textbf q}^\star_{\pi\pi} \big\rangle  =\sum_{\ell,m_\ell} \sqrt{4 \pi} \, Y_{\ell m_\ell}(\hat{\textbf q}^\star_{\pi\pi})  \,     \big| P_{\pi\pi}, \ell ,m_\ell \big\rangle,
\end{equation}
and this is the definition used in defining the transition amplitude given in Eq.~(\ref{eq:infinite_volume_amp}).

\section{Contamination from $\ell\geq3$ partial waves \label{sec:F_wave}}
 
Although in this first calculation we have not explicitly determined the contribution of $\ell\geq 3$ partial waves
\footnote{Here $\ell$ denotes the orbital angular momentum of the $\pi\pi$ state, which is equal to the total angular momentum, $J$, of the $\pi\gamma^\star$ state.}
, we can give an analytic expression that describes how they appear in the relation between finite and infinite-volume quantities. As demonstrated in Refs.~\cite{Meyer:2012wk, Briceno:2015csa, Briceno:2014uqa}, due to the reduction of rotational symmetry in a cubic volume, transition amplitudes involving different partial waves appear together in finite-volume irreps, leading to $ \mathcal{R}$ in Eq.~(\ref{eq:Rintro}) being a matrix in $\ell$-space. Expanding the denominator about $E_{\textbf{2}}$ we find
\begin{align}
\mathcal{R}(E_{\textbf{2}}, \textbf P) &=F(P,L)\frac{{\rm adj}[\mathbb{M}]}{\text{tr}\left[ \text{adj}[{\mathbb{M}}]\frac{\partial {\mathbb{M}}}{\partial E_2}\right]}  \mathcal M(P)^{-1}
, \label{B1}
\end{align}
where
\begin{align}
\mathbb{M}= \mathcal M(P)^{-1}+ F(P,L),
 \end{align}
is purely real, and ${\rm adj}[\mathbb{M}]$ is its adjoint (or adjugate). Since we are simply interested in the mixing due to the lowest-lying higher partial wave above $\ell=1$, we will restrict our attention to the scenario where we are dealing with two-dimensional matrices with $\ell=1,3$. At low energy we are justified in neglecting the $\ell=3$ contribution to the elastic scattering amplitude (see Ref.~\cite{Dudek:2012xn}) so,
\begin{align}
\mathcal{M}= 
\begin{bmatrix} 
			\mathcal{M}_{\ell=1} & 0\\
			0 & \mathcal{M}_{\ell=3}
		\end{bmatrix}
\approx
\begin{bmatrix} 
			\mathcal{M}_{\ell=1} & 0\\
			0 & 0
		\end{bmatrix},
 \end{align}
but the finite-volume function $F$ is generally not diagonal in angular momentum, so
\begin{align}
F= 
\begin{bmatrix} 
			F_{11} & F_{13}\\
			F_{13} & F_{33}
		\end{bmatrix}.
 \end{align}
The adjoint of $\mathbb{M}$ is easily evaluated,
\begin{eqnarray}
\mathrm{adj}[{\mathbb{M}}]=
\left(
\begin{array}{cc}
~~{\mathbb{M}}_{22}
 & 
-{\mathbb{M}}_{12} \\
- {\mathbb{M}}_{21}&
~~{\mathbb{M}}_{11}
  \\
\end{array}
\right),
\label{eq:2Dimadj}
\end{eqnarray}
and in the limit that the $\ell=3$ elastic scattering amplitude is zero, the spectrum satisfies $\mathcal{M}^{-1}_{\ell=1}=-F_{11}$ and we obtain
\begin{align}
\text{tr}\left[ \text{adj}[{\mathbb{M}}] \, \frac{\partial {\mathbb{M}}}{\partial E_2}\right] 
=
\mathcal{M}^{-1}_{\ell=3}
\times
\frac{\partial 
}{\partial E_2}
\big(\mathcal{M}^{-1}_{\ell=1}+F_{11}  \big)
\end{align}
and
\begin{align}
F(P,L)
\,{\mathrm{adj}[\mathbb{M}]}
\, \mathcal M(P)^{-1}
&=\nn\\
&\hspace{-1.3cm}
\begin{bmatrix} 
			{F_{11}}\, {\mathcal M_{\ell=1}^{-1}} & 
			{F_{13}}\, {\mathcal M_{\ell=1}^{-1}} \\
			 {F_{13}}\, {\mathcal M_{\ell=1}^{-1}}  & 
			 -F_{13}^2
		\end{bmatrix}
	\times
	\mathcal{M}^{-1}_{\ell=3},
\end{align}
and $\mathcal{M}^{-1}_{\ell=3}$ cancels in the ratio in Eq.~\ref{B1}.

The end result is that allowing a non-zero $\ell=3$ transition amplitude but with negligible $\ell=3$ elastic scattering amplitude means that Eq.~(\ref{eq:master_equation}) is given by
\begin{align}
 \label{eq:F_wave_contamination}
\big|{\langle  \textbf{1};L|\mathcal{J}^\mu(0)| \textbf{2};L\rangle} \big|
&=\frac{1}{L^3} \frac{1}{\sqrt{2E_\textbf{1}}}
\\
&\hspace{-2cm}
\times
\sqrt{
\frac{
c_1\left(\mathcal{H}_{\ell=1}^{\mu}
\right)^2
+
c_2\mathcal{H}_{\ell=1}^{\mu}\mathcal{H}_{\ell=3}^{\mu}
+
c_3\left(\mathcal{H}_{\ell=3}^{\mu}
\right)^2
}
{
{\frac{\partial 
}{\partial E_2}
(\mathcal{M}^{-1}_{\ell=1}+F_{11})
}}
}
,
~\end{align}
where $c_1={F_{11}}\, {\mathcal M_{\ell=1}^{-1}}$, $c_2={F_{13}}\, {\mathcal M_{\ell=1}^{-1}}$ and $c_3=-(F_{13})^2$. Equivalently, using the quantization condition, one can write these as 
\begin{align}
c_1&=-(F_{11})^2 \propto -(\cot\phi_{1}+i)^2=-\frac{e^{i2\phi_{1}}}{\sin^2\phi_{1}} ,\\
c_2&\propto -(\cot\phi_{1}+i)\cot\phi_{13}=-\frac{e^{i\phi_{1}}\cot\phi_{13}}{\sin\phi_{1}} ,\\
c_3&\propto -\cot^2\phi_{13} .
\end{align}
Note, $c_1$ and $c_2$ are in general complex while $c_3$ is real. This is consistent with the fact that the term inside of the square root in Eq.~(\ref{eq:F_wave_contamination}) must be real. According to Watson's theorem $\mathcal{H}_{\ell=1}^{\mu}\propto e^{i\delta_1}=e^{-i \phi_1}$, while $\mathcal{H}_{\ell=3}^{\mu}\propto e^{i\delta_3}=1$ in our approximation of no elastic scattering in $\ell=3$. 

As an example, for the $T^-_1$ irrep, one finds~\cite{Luscher:1990ux}
\begin{align}
\cot \phi_{1} &= \cot\phi^{\textbf{0}}_{00} \,\\
\cot \phi_{13} &= \frac{4}{\sqrt{21}}\cot\phi^{\textbf{0}}_{40} \,,
\end{align}
where the pseudophases, $\phi^\textbf{P}_{lm}$, are those defined in Eq.~(\ref{eq:philm}).

In order to estimate the contribution due to the ${\ell \geq 3}$ transition amplitudes, one could perform calculations of three-point functions using irreps where the $\pi\pi$ state couples to $\ell=3$ but not to the $\ell=1$ partial wave, for example the $[001]\,B_1$ and $[001]\,B_2$ irreps.


\section{Symmetry factor and identical particles \label{sec:symmetry_factors}}

In Eq.~(\ref{eq:LLfactorP}) we gave the definition of the LL-factor for distinguishable particles. In general one should write the LL-factor as 
\begin{align}
\frac{2 E_\pi}{\RLL}
&=
\frac{1}{\xi}
32\pi \, \frac{E_\pi \Epipi}{q^{\star}_{\pi\pi}}\,
\big(\delta_1'+r\phi' \big)
\label{eq:LLfactorP_dist}
\end{align}
where $\xi$ is the `symmetry factor', which is equal to 1/2 if the particles are indistinguishable and 1 otherwise. For the system of interest the interpretation of this factor is a subtle one. Given that the $\pi\pi\to\pi\gamma^\star$ transition can only take place if the initial $\pi\pi$ system is in a parity-odd state, with the bosonic nature of the $\pi$ one is lead to believe that the initial state must be composed of distinguishable particles, e.g. $\pi^+\pi^0$, and consequently this symmetry factor should not appear. In the limit of perfect isospin symmetry, which we have in this calculation, the eigenstates of the Hamiltonian are of definite isospin. Therefore one has a choice whether to evaluate matrix elements featuring $|\pi^+\pi^0,\ell=1\rangle$ or those of definite isospin, ${|\pi\pi,I=1, m_I=+1,\ell=1\rangle}$ given by ${(|\pi^+\pi^0,\ell=1\rangle-|\pi^0\pi^+,\ell=1\rangle)/\sqrt{2}}$. The presence of the symmetry factor differs depending on this choice. For example, for the elastic scattering amplitude and transition amplitude the choices are related via
\begin{align}
\mathcal M_{\ell=1,I=1,m_I=+1 }&=2~\mathcal M_{\ell=1,\pi^+\pi^0 },\nn\\
{\mathcal{H}_{\pi\pi,\pi\gamma^\star,I=1,m_I=+1}^{\mu} }&=\sqrt{2}~{\mathcal{H}_{\pi^+\pi^0,\pi\gamma^\star}^{\mu} } \,.
\end{align}

The definition of the finite-volume matrix element, Eq.~(\ref{eq:master_equation}), can be seen to be independent of the symmetry factor,
\begin{align}
\big|{\langle  \textbf{1};L|\mathcal{J}^\mu(0)| \textbf{2};L\rangle}\big|
&=\frac{\sqrt{\mathcal{H}_{1,2}^{\mu}~\mathcal{R}~\mathcal{H}_{2,1}^{\mu}}}{L^3~\sqrt{2E_1}}\nn\\
&\propto
\sqrt{\xi^{-1/2}~\xi~\xi^{-1/2}}= 1.
~\end{align}

By not introducing the symmetry factor in Eq.~(\ref{eq:LLfactorP}), we are determining the amplitudes using the $\big|\pi^+\pi^0,\ell=1 \big\rangle$ basis for asymptotic states. Doing so allows us to more easily compare phenomenological extractions from experimental data where asymptotic states are not constructed in the isospin basis.  


\section{Lorentz covariant decompositions of the matrix elements \label{sec:covariant_relation}}

In this appendix we show that the decomposition of the $P$-wave matrix element in Eq.~\ref{eq:decomp} is equivalent to another common decomposition. The Lorentz invariant transition amplitude may be obtained by contracting the matrix element of the electromagnetic current with the polarization vector of the photon, $\epsilon^\mu(q,\lambda_\gamma)$, where $\lambda_\gamma$ is the helicity of the photon,
\begin{align}
{\langle \pi\pi|{\mathcal{J}}^{\mu}(0)| \pi\rangle}\epsilon_\mu(q,\lambda_\gamma)=
\mathcal{M}_{\lambda_\gamma}.
\end{align}

A common decomposition for the  ${\gamma^\star(q, \lambda_\gamma)\, \pi(p_1) \to \pi(p_2)\, \pi(p_3)}$ amplitude, not projected into any particular partial wave, is
\begin{equation}
\mathcal{M}_{\lambda_\gamma} = \epsilon_{\mu \nu \rho \sigma}\,  \epsilon^\mu(q,\lambda_\gamma) p_1^\nu \, p_2^\rho \, p_3^\sigma \; T(s,t,Q^2) \label{pipi_invar}
\end{equation}
where the invariant amplitude, $T(s,t,Q^2)$, is a function of $s = (q+p_1)^2$, $t = (p_1 - p_2)^2 $, and the virtuality of the photon, $Q^2$.

In our case we are interested in the amplitude for the $P$-wave, which can be obtained in the standard way~\cite{Jacob:1959at, Danilkin:2014cra} by partial-wave expanding $\mathcal{M}_{\lambda_\gamma}$,
\begin{equation}
\mathcal{M}_{\lambda_\gamma}=\sum_{J=1,3,\ldots} (2J+1) \, d^{(J)}_{\lambda_\gamma, 0}(\theta) \, A_{J; \lambda_\gamma}(s,Q^2),
\end{equation}
where we have chosen the scattering plane to have $\phi=0$, and where $d^{(J)}_{\lambda_\gamma, 0}(\theta)$ are the reduced Wigner $d$-functions. Enforcing parity conservation ensures that $A_{J; 0}(s,Q^2)=0$ and $A_{J; -1}(s,Q^2)=-A_{J; 1}(s,Q^2)$. The contribution of the $P$-wave can be isolated,
\begin{equation}
\mathcal{M}_{\lambda_\gamma}=-\frac{3}{\sqrt{2}} \lambda_\gamma \, \sin \theta \; A_{1; \lambda_\gamma}(s,Q^2) +\cdots,
\label{eq:MatPwave}
\end{equation}
with the ellipses denoting the higher partial-wave contributions. 

The decomposition in Eq.~(\ref{pipi_invar}) is most easily investigated in the cm-frame. If we let the incoming states have momenta lying along the $\hat{z}$-axis and the outgoing momenta in the $\hat{x}\hat{z}$-plane,
\begin{align*}
q^\mu = \big( E_\gamma, 0, 0, q \big) 	\quad& p_2^\mu = \big( E', k\sin \theta , 0, k \cos \theta \big) \\
p_1^\mu = \big( E_1, 0, 0, -q \big) 	\quad& p_3^\mu = \big( E', -k\sin \theta , 0, -k \cos \theta \big), 
\end{align*}
with the photon polarization vector being ${\epsilon^\mu(q, \lambda_\gamma = \pm 1) = \mp \tfrac{1}{\sqrt{2}} \big( 0, 1, \pm i , 0 \big)}$. It follows that $\mathcal{M}_{\lambda_\gamma} =	-\sqrt{2} i \,  T(s,t,Q^2)\, k q E' \sin \theta $, and the presence of a single factor of $\sin \theta$, as is the case for the $P$-wave in Eq.~(\ref{eq:MatPwave}), indicates that the $P$-wave part of the amplitude must lack any further $t$-dependence in $T(s,t,Q^2)$. We also note that $\mathcal{M}_{\lambda_\gamma}$ contains explicitly the factors $q$ and $k$ which describe the $P$-wave threshold behavior in the initial and final states. In light of this we can write an invariant decomposition capable of describing the $P$-wave as
\begin{equation}
\mathcal{M}_{\lambda_\gamma}^{\mathrm{[1]}} = \epsilon_{\mu \nu \rho \sigma}\,  \epsilon^\mu(q,\lambda_\gamma) p_1^\nu \, p_2^\rho \, p_3^\sigma \; T_1(s,Q^2) \label{pipi_P_invar},
\end{equation}
where $T_1(s,Q^2)$ should not have the $\propto k$, $\propto q$ threshold behavior and where the superscript ${\rm ``[1]"}$ denotes this is the first of two decompositions we are relating.

We are now in a place to reconcile this decomposition with the one used through this work, Eq.~\ref{eq:decomp}, which we rewrite here using the variables defined in this appendix,
\begin{equation}
\mathcal{M}_{\lambda_\gamma}^{[2]} =\epsilon_{\mu \nu \rho \sigma} \epsilon^\mu(q,\lambda_\gamma)\,  p_1^\nu\,  \epsilon^{\rho*}(P,\lambda) P^\sigma\; \mathcal A(s,Q^2) , \label{rho_pi_gamma}
\end{equation}
where $P^\sigma=(p_2+p_3)^\sigma$ and $\epsilon^{\rho*}(P,\lambda)$ is the polarization vector of the $\pi\pi$ system which has been projected in a \mbox{$P$-wave} with helicity $\lambda$. In this appendix we are considering the time-reversed process, $\gamma^\star \, \pi \to \pi\, \pi$ which explains the presence of the complex conjugate of the $\pi\pi$ polarization vector. 

The claim is that Eq.~\ref{rho_pi_gamma} is equivalent to Eq.~\ref{pipi_P_invar}, after the $\pi\pi$ state appearing in the latter has been projected in a \mbox{$P$-wave}. To show this, we begin by constructing a $\pi\pi$ helicity state in the cm-frame,
\begin{equation}
\big| |\textbf{k}|; J=1, \lambda \big\rangle = \int \! d\hat{\mathbf{k}} \, \frac{Y_{1\lambda}(\hat{\mathbf{k}})}{\sqrt{4\pi}}\, \big| \pi(\textbf{k}) \pi(-\textbf{k})\big\rangle,
\end{equation}
which we can boost to a frame having momentum $\textbf{P}$ by first boosting the system along the $\hat{z}$-axis and then performing a rotation to the axis of the momentum
\begin{equation}
\big| \textbf{P}; |\textbf{k}|; J=1, \lambda \big\rangle = U[R(\hat{P})]\, U[Z_P]\,  \big| |\textbf{k}|; J=1, \lambda \big\rangle.
\end{equation}

The $\hat{z}$-axis boost acting on four-vectors can be expressed as
\begin{equation}
	\big[ Z_P \big]^\mu_\nu = 
		\begin{bmatrix} 
			\gamma & 0 & 0 & \beta \gamma \\
			0 & 1 & 0 & 0 \\
			0 & 0 & 1 & 0 \\
			\beta \gamma & 0 & 0 & \gamma	
		\end{bmatrix}
		= \frac{1}{2 \omega_\pi}
		\begin{bmatrix} 
			\Epipi & 0 & 0 & |\textbf{P}|\\
			0 & 1 & 0 & 0 \\
			0 & 0 & 1 & 0 \\
			|\textbf{P}| & 0 & 0 & \Epipi	
		\end{bmatrix}
\end{equation}
since $\gamma = \frac{\Epipi}{2\omega_\pi}$ and $\beta \gamma = \frac{|\textbf{P}|}{2 \omega_\pi}$ where $\omega_\pi = \sqrt{m_\pi^2 + k^2}$. Then the action of the boost on $k^\mu = \big(\omega_\pi, \textbf{k}\big)$ and ${\bar{k}^\mu = \big(\omega_\pi , -\textbf{k} \big)}$ is
\begin{align}
k'^\mu &= \big[ Z_P \big]^\mu_\nu \, k^\nu = 
	\begin{bmatrix} \tfrac{\Epipi}{2} + \tfrac{|\textbf{P}|}{2\omega_\pi} k_z \\ k_x \\ k_y \\ \tfrac{|\textbf{P}|}{2} + \tfrac{\Epipi}{2\omega_\pi} k_z \end{bmatrix},
	\quad \quad\quad\nn\\
\bar{k}'^\mu &= \big[ Z_P \big]^\mu_\nu \, \bar{k}^\nu = 
	\begin{bmatrix} \tfrac{\Epipi}{2} - \tfrac{|\textbf{P}|}{2\omega_\pi} k_z \\ -k_x \\ -k_y \\ \tfrac{|\textbf{P}|}{2} - \tfrac{\Epipi}{2\omega_\pi} k_z \end{bmatrix},	
\end{align}
and as expected, $\bar{k}'^\mu = P^\mu - k'^\mu$. It follows that 
\begin{equation}
\big| \textbf{P}; |\textbf{k}|; J=1, \lambda \big\rangle = \int \! d\hat{\mathbf{k}} \, \frac{Y_{1\lambda}(\hat{\mathbf{k}})}{\sqrt{4\pi}}\, \big| \pi(R\textbf{k}') \pi( \textbf{P} - R\textbf{k}') \big\rangle.
\end{equation}

We can write the matrix element
\begin{align}
 \big\langle \textbf{P}; |\textbf{k}|; J=1, \lambda \big| {\mathcal{J}}^{\mu}(0) \big| \gamma(q, \lambda_\gamma)\, \pi(p_1) \big\rangle &=\nn\\
 &\hspace{-5.5cm} 
 \int \! d\hat{\mathbf{k}} \, \frac{Y_{1\lambda}^*(  \hat{\mathbf{k}})   }{\sqrt{4\pi}}\,  \big\langle \pi(R\textbf{k}') \pi( \textbf{P} - R\textbf{k}') \big| {\mathcal{J}}^{\mu}(0) \big| \gamma(q, \lambda_\gamma)\, \pi(p_1) \big\rangle,
\end{align}
and substituting in the decomposition in Eq.~(\ref{pipi_P_invar}) we have
\begin{align}
\mathcal{M}_{\lambda_\gamma}^{[1]}  
 &= T_1(s,Q^2)\, \epsilon_{\mu\nu\rho\sigma} \, \epsilon^\mu(q,\lambda_\gamma) \, p_1^\nu \, P^\sigma \nn\\
& \times\; \int \! d\hat{\mathbf{k}} \, \frac{Y_{1\lambda}^*(\hat{\mathbf{k}})}{\sqrt{4\pi}}\,\big(R k')^\rho\,
 \end{align}
which will be equivalent to Eq.~(\ref{rho_pi_gamma}) if $\int \! d\hat{\mathbf{k}} \, Y_{1\lambda}^*(\hat{\mathbf{k}})\,\big(R k')^\rho$ transforms in the same way as $\epsilon^{\rho*}(P, \lambda)$. Since the rotation can be factored out of the integral, and since $\epsilon^\rho(RP_z, \lambda) = \big[ R \big]^\rho_\sigma \epsilon^\sigma(P_z, \lambda)$, it follows that we just need to show that $X^\sigma(\lambda) = \int \! d\hat{\mathbf{k}} \, Y_{1\lambda}(\hat{\mathbf{k}})\, k'^\sigma$ transforms like $\epsilon^{\sigma}(P_z, \lambda)$. First we establish that $P_\mu X^\mu = 0$,
\begin{align}
	P_\mu X^\mu 
 	&= \int \! d\hat{\mathbf{k}} \, Y_{1\lambda}(\hat{\mathbf{k}})\, \left[ \frac{\Epipi^2}{2}  - \frac{P^2}{2} \right] = 0,
\end{align}
and then we may check that the $\lambda=\pm 1$ components are what is expected, e.g.,
\begin{align}
	X^\sigma(\lambda = +1) &= \int \! d\hat{\mathbf{k}} \, Y_{1,+1}(\hat{\mathbf{k}})\, k'^\sigma \nn\\
	&= -\sqrt{\frac{4\pi}{3}} |\textbf{k}| \frac{1}{\sqrt{2}} \begin{bmatrix}
		0 \\ 1 \\ i \\ 0
	\end{bmatrix}\nn\\
	&= \sqrt{\frac{4\pi}{3}} |\textbf{k}| \epsilon^\sigma(P_z,\lambda=+1),
\end{align}
and indeed the forms are equivalent. Note the presence of a factor of $|\textbf{k}| = k$ above, which suggests that $\mathcal A \sim k \, T_1$. Recalling that $T_1$ does not have the threshold factor for the final state $\pi \pi$, we see that in the case that the $\rho$ is unstable into $\pi \pi$, the quantity $\mathcal A$ should behave like $k$ around the $\pi \pi$ threshold.


\section{Cross-sections \label{app:cross_section}}

In this appendix we derive the relation given in Eq.~(\ref{eq:cross_section}) for the cross-section with a real photon. We begin with the standard definition of the differential cross section,
\begin{align}
\label{eq:diff_cs}
\frac{d\sigma}{d\Omega}\big(\pi^+\gamma\to(\pi^+\pi^0)_{\lambda} \big)   =   \frac{1}{64\pi^2}
\frac{{q}^\star_{f}}{{q}^\star_{i}}\frac{1}{E_{\pi\pi}^{\star2}} e^2  \,  \Big|\mathcal{M}^{[2]}_\lambda   \Big|^2 
\end{align}
where $\lambda$ is the helicity of the final state and $\mathcal{M}^{[2]}_\lambda$ has been defined in Eq.~(\ref{rho_pi_gamma}). To obtain the total cross-section, we average over the initial photon helicity and sum over the helicity of the final $\pi\pi$ state, and this gives
\begin{align}
\sigma(\pi^+\gamma\to\pi^+\pi^0)\equiv
\frac{1}{2}\sum_{\lambda, \lambda_\gamma}
\int \!\!d\Omega\, \frac{d\sigma}{d\Omega} \big(\pi^+\gamma\to(\pi^+\pi^0)_{\lambda}\big),
\end{align}
which is proportional to
\begin{align}
\frac{1}{2}&\sum_{\lambda, \lambda_\gamma}
\Big|\mathcal{M}^{[2]}_\lambda  \Big|^2\nn\\
&
= \tfrac{1}{2} \big|\mathcal A(\Epipi^{\star},0)\big|  \sum_{\lambda, \lambda_\gamma} 
\epsilon_{\mu \nu \rho \sigma} \epsilon^\mu(q,\lambda_\gamma)\,  p_1^\nu\,  \epsilon^{\rho*}(P,\lambda) P^\sigma\nn\\
&\hspace{3cm}\times \epsilon_{\bar \mu \bar \nu \bar \rho \bar \sigma} \epsilon^{\bar \mu*}(q,\lambda_\gamma)\,  p_1^{\bar \nu}\,  \epsilon^{\bar \rho}(P,\lambda) P^{\bar \sigma}\nn\\
&
=\tfrac{1}{2} \big|\mathcal A(\Epipi^{\star},0)\big| \epsilon_{\mu \nu \rho \sigma} \epsilon_{\bar \mu \bar \nu \bar \rho \bar \sigma} \big(-g^{\mu\bar\mu}\big) \left(-g^{\rho\bar\rho}+ \frac{P^{ \rho}P^{\bar \rho}}{E_{\pi\pi}^{\star2}} \right)\nn\\
&\hspace{3.5cm}\times
 p_1^\nu\, p_1^{\bar \nu} \,  P^\sigma  \,  P^{\bar \sigma}\nn\\
 &
=\tfrac{1}{2} \big|\mathcal A(\Epipi^{\star},0)\big|    \epsilon_{\mu  \rho \nu \sigma} \epsilon^{ \mu \rho ~~}_{~~ \bar \nu \bar \sigma} ~
 p_1^\nu\, p_1^{\bar \nu} \,  P^\sigma  \,  P^{\bar \sigma},
\end{align}
and evaluating the tensor contraction and writing in terms of cm-frame quantities this becomes $\big|\mathcal A(\Epipi^{\star},0)\big|^2\,   E_{\pi\pi}^{\star2} \,{q}^{\star2}_{i} $ and for the cross-section we have
\begin{align}
\sigma(\pi^+\gamma\to\pi^+\pi^0)&=
\frac{e^2}{4\pi}
\frac{{q}^\star_{f}   \, {q}^\star_{i}}{m_{\pi}^2}  
\big|\mathcal A(\Epipi^{\star 2},0) \big|^2.
\end{align}

The cross section can be expressed in terms of the form factor, $F(E_{\pi\pi}^\star, Q^2)$, using Eqs.~\ref{eq:AtoF} and~\ref{eq:KF} as, 
\begin{align}
\label{eq:cross_section_param2}
\sigma(\pi^+\gamma\to\pi^+\pi^0)
&= 16 \pi \, \alpha \, \frac{q^\star_{i}}{m_{\pi}^2} \big| F(E_{\pi\pi}^\star, Q^2) \big|^2 \, \frac{\sin^2 \!\delta_1(E_{\pi\pi}^\star)}{\Gamma(E_{\pi\pi}^\star)},
\end{align}
from which it is easy to find the peak cross-section by evaluating when $\delta_1 = 90^\circ$. Comparing to the expression given in Ref.~\cite{Capraro:1987rp}, where the cm energy $E_{\pi\pi}^\star$ has been approximated by the real part of the $\rho$ mass, we find a definition of the radiative decay width of $\rho^+\to\pi^+\gamma$ in terms of the form factor,
\begin{align}
\label{eq:decay_rate}
\Gamma(\rho^+\to\pi^+\gamma)=\alpha \frac{4}{3}\frac{{q}^\star_{i}}{m_\pi^2}~ \big|F(m_\rho,0)\big|^2.
\end{align}


 
\bibliography{bibi}

\end{document}